\documentclass[a4paper,usenatbib]{mnras}

\usepackage{mathptmx}

\usepackage[T1]{fontenc}
\usepackage{ae,aecompl}

\usepackage{graphicx}	
\usepackage{amsmath}	
\usepackage{amssymb}	
\usepackage{lmodern}
\usepackage[english]{babel}
\usepackage{array}
\usepackage{mathrsfs}
\usepackage{stackrel}
\usepackage{wasysym}
\usepackage{esint}
\usepackage{dsfont}

\usepackage{color}

\title[Improving Fisher matrix forecasts for galaxy surveys]{Improving Fisher matrix forecasts for galaxy surveys: window function, bin cross-correlation, and bin redshift uncertainty}

\author[A. Bailoni, A. Spurio Mancini and L. Amendola]{
Alberto Bailoni,$^{1}$\thanks{bailoni@thphys.uni-heidelberg.de}
Alessio Spurio Mancini,$^{1}$\thanks{spuriomancini@thphys.uni-heidelberg.de}
and Luca Amendola$^{1}$\thanks{l.amendola@thphys.uni-heidelberg.de}
\\
$^{1}$Institut f\"ur Theoretische Physik, Ruprecht-Karls-Universit\"at Heidelberg,
Philosophenweg 16, 69120 Heidelberg, Germany
}

\date{Accepted XXX. Received YYY; in original form ZZZ}

\pubyear{2016}

\begin{document}
\label{firstpage}
\pagerange{\pageref{firstpage}--\pageref{lastpage}}
\maketitle

\begin{abstract}
The Fisher matrix is a widely used tool to forecast the performance
of future experiments and approximate the likelihood of large data
sets. Most of the forecasts for cosmological parameters in galaxy
clustering studies rely on the Fisher matrix approach for large-scale
experiments like DES, Euclid, or SKA. Here we improve upon the standard
method by taking into account three effects: the finite window function,
the correlation between redshift bins, and the uncertainty on the
bin redshift. The first two effects are negligible only in the limit
of infinite surveys. The third effect, on the contrary, is negligible
for infinitely small bins. Here we show how to take into account these
effects and what the impact on forecasts of a Euclid-type experiment
will be. The main result of this article is that the windowing and
the bin cross-correlation induce a considerable change in the forecasted
errors, of the order of 10-30\% for most cosmological parameters,
while the redshift bin uncertainty can be neglected for bins smaller
than $\Delta z=0.1$ roughly.
\end{abstract}



\onecolumn
\section{Introduction}

An important task of cosmology is to study the composition and evolution
of the universe. Different models have been introduced, which rely
on cosmological parameters like the Hubble constant, the primordial
scalar spectral index, and matter abundances. From a theoretical perspective
it is crucial to distinguish between different models and determine
which of them provide the closest approximations to the observed data.
In a few years, large-scale surveys like Euclid \citep{euclidredbook},
DESI \citep{DESI}, HETDEX \citep{0806.0183}, eBOSS \citep{eBOSSonline},
BigBOSS \citep{schlegel2011bigboss}, DES \citep{collaboration2005dark},
Pan-STARRS \citep{kaiser2002pan}, LSST \citep{abell2009lsst} and SKA
\citep{yahya2015cosmological} will provide huge datasets containing
information on galaxy positions at high redshift. The data will allow
us to study how the observed clustering of galaxies evolves over time,
and how the gravitational lensing is generated by large-scale structures
in the universe. The data will be mainly encoded in 3D or angular
power spectra and in higher order moments, since these descriptors
are usually the direct outcome of cosmological theories. The data
will then be combined with the cosmic microwave background, in particular
with the Planck satellite \citep{2015arXiv150201589P}.

The first attempt to characterize large scale structure and measure the power spectrum from data was presented in \cite{yu1969superclusters} and \cite{groth1977statistical}. In \cite{baumgart1991fourier}, the result was generalized for redshift surveys and full 3-dimensional galaxy positions. In 1994, \cite{FKP} provided an estimate of the power spectrum and, in a following pioneering paper, \cite{tegmark1997measure} introduced an optimal method for estimating the power spectrum
based on Bayesian statistics and the Fisher matrix formalism \citep{fisher1935logic}.

Given the measured power spectrum, a maximum likelihood analysis yields then the best estimate of the parameters characterizing the theoretical power spectrum and the first attempts to apply this method at linear scales were made by \cite{fisher1994spherical} and \cite{heavens1995spherical}. To formalize the theoretical power spectrum, the most commonly employed methods are the following two: the first uses real space coordinates, whereas the second expresses the power spectrum in angular coordinates
and redshift space, so that it is directly related to the observable angular correlation function. The latter method is based on a spherical harmonic decomposition of the projected density in redshift shells, as shown in \cite{heavens1995spherical} (for recent applications, see \cite{montanari2012new,montanari2015measuring,raccanelli2016doppler,bonaldi2016ska}). In this approach, one has the important advantage that large-scale relativistic effects, e.g. lensing, are straightforwardly included, as well as correlations between different shells. On the other hand, the 3D nature of the correlation is compressed into a set of 2D projections and some information is then lost. This information is completely recovered
in the limit of an infinite redshift-resolution along the line of sight, when the two power spectrum formalizations become equivalent.

Even before performing large-scale surveys, when data are not available yet,
 it is important to know which set-up will allow us to optimally distinguish between theoretical models with different parameters. Given the specifications of the survey and the model, we can compute the probability distributions of the parameters that define the model by using Bayesian statistics and the Fisher matrix formalism. Such predictions, called \emph{forecasts}, are crucial for setting up a survey and defining the models to focus on. If the survey is still in the design phase, then several survey-specific parameters can also be included in the forecast, e.g. the density of galaxies, redshift errors, the survey's depth and sky coverage.

Analyzing a large dataset of 3D galaxy positions in an expanding universe is time-consuming. The standard procedure for a galaxy redshift
survey is to divide the survey space in $N$ redshift bins, so that
each bin represents data from a different cosmological epoch. This
can be formalized by introducing the concept of a top-hat survey-bin
window function $W_{i}(\mathbf{x})$, such that $\int W_{i}(\mathbf{x})d^{3}x=V_{i}$
and $W_{i}(\mathbf{x})$ is equal to one inside the $i$-th survey-bin
$V_{i}$ and zero otherwise. Most of the forecast studies based on a real space description of the power spectrum (e.g. \cite{seo03,2005MNRAS.357..429A,wang2006dark,albrecht2009findings,di2012simultaneous,wang2012robust,xu2014forecasts}) typically imply several assumptions, four of which are spelled out now. The first assumption is that the
power spectrum computed for a finite redshift bin volume is not significantly
different from the power spectrum computed with respect to the full
sky, i.e. the window function effect on the matter power spectrum
can be neglected. The second assumption is that the bin cross-correlation
spectra can be ignored, i.e. only correlations between galaxies from
the same redshift bins are considered. Thirdly, the cross-correlations
among different $k$-modes, arising because of the finiteness of the
survey volume, are also neglected. Whereas for an infinite survey
these assumptions would hold, they have never been properly evaluated
for a finite survey\footnote{Note that for a power spectrum expressed in redshift space, cross-correlation spectra and window
function effects have already been taken into account with lensing effects included \citep{montanari2015measuring}.}. The fourth assumption is that one can assign to the $z$-dependent
functions, e.g. the growth rate, the growth function, the bias, the
Hubble function, etc., a precise redshift value, usually taken to
be the median redshift of the bin. This is an approximation that will
also be discussed later on.

In this article, we focus on forecast studies for galaxy clustering surveys and test how well the
above assumptions hold for future surveys, in particular the Euclid
survey. Our first contribution is the formalization of the effect
of the window function, the bin cross-correlations, and the bin redshift
uncertainty; although we focus on forecast studies, this theoretical description can also contribute to parameter estimation when data are available. As second contribution, we quantify the effects of the related assumptions by
comparing the forecast confidence regions for the cosmological parameters computed
with and without each of the mentioned assumptions. Given the specifications
of the planned Euclid survey, we investigate how these effects depend
on the number of redshift bins. Then, we study how to introduce the
Alcock-Paczynski effect \citep{Alcock:1979mp}. We compute the windowed
and the cross-correlation spectra using an FFT algorithm \citep{press2007numerical}.
Its output is further processed using an optimized Fisher matrix implementation,
which we make publicly available.\footnote{\url{https://github.com/abailoni/fisher-correlations}}
The main result of this article is that the window function effect
increases the uncertainties because it flattens out the signal, while
numerically we find that the bin cross-correlation reduces them 
(see Sec. \ref{sub:Testing-the-assumptions}); the combined effect
induces a sizable \emph{increase} in the forecasted errors, of the
order of 10-30\% for most cosmological parameters. Finally, the bin
redshift uncertainty always increases the overall uncertainty, but
it is important only for bins larger than $\Delta z\approx0.1$.

The structure of the article is the following. In section \ref{sec:The-observed-matter},
we briefly introduce the relevant theoretical concepts related to
the linear matter power spectrum and redshift distortions, focusing
on the $\Lambda$CDM model. In section \ref{sec:Correlations-among-bins},
we overview the Fisher matrix approach, the effects of the window
function, the bin cross-correlations, and the bin redshift uncertainty.
In section \ref{sec:computing-obs-spectr}, we describe how to include
the effects of the redshift distortions in the convolved and cross-correlation
spectra and tune the parameters of the radial FFT algorithm based
on analytical power spectra. In section \ref{sec:Results-and-discussion},
we discuss the results we obtained. Section \ref{sec:Conclusions}
concludes the article.

\section{The observed matter power spectrum}\label{sec:The-observed-matter}

In this section, we will briefly introduce the theoretical concepts
regarding the linear matter power spectrum, the relevant cosmological
quantities and the redshift distortions. Let us first define the ensemble-averaged
power spectrum $P(\mathbf{k})$ of a survey in a volume $V$:
\begin{equation}
V\left\langle \delta_{\mathbf{k}}\delta_{\mathbf{k'}}^{*}\right\rangle \equiv\frac{(2\pi)^{3}}{V}P(\mathbf{k})\delta_{D}(\mathbf{k-k'}),\label{eq:def_pow_spectr}
\end{equation}
where $\delta_{\mathbf{k}}$ are the Fourier coefficients of the density
contrast $\delta(\mathbf{x})$ and $\delta_{D}(\mathbf{k})$ is the
Dirac delta.  The shell volume $V$ comes from the normalization convention for the 3-dimensional Fourier transform of a generic function $f(\mathbf{x})$:

\begin{align}
f(\mathbf{x}) = \frac{V}{(2 \pi)^3}\int f_{\mathbf{k}} e^{i \mathbf{k} \cdot \mathbf{x}} \mathrm{d}^3 k, \\
f_{\mathbf{k}} = \frac{1}{V} \int f(\mathbf{x})  e^{- i \mathbf{k} \cdot \mathbf{x}} \mathrm{d}^3 k.
\end{align}

The data power spectrum is a single sampling of the distribution.
Nevertheless, when an average on the theoretical prediction is considered,
an ensemble average and a sample one are equivalent, assuming that
the survey is a fair example of the universe \citep{Amendola2010}.
Given Eq. (\ref{eq:def_pow_spectr}), it follows that for equal wave
vectors the power spectrum is $P(\mathbf{k})=V\langle\delta_{\mathbf{k}}\delta_{\mathbf{-k}}\rangle$.\footnote{Note that since $\delta(\mathbf{x})$ is real, then $\delta_{\mathbf{k}}^{*}=\delta_{-\mathbf{k}}$.}
If the field is then assumed to be isotropic, the power spectrum will
depend only on the modulus $k$.

We assume a flat, homogeneous and isotropic metric. Then, neglecting the small contribution given by radiation, the Hubble
parameter $H(z)$ is
\begin{equation}
H(z)=H_{0}\sqrt{\Omega_{m}^{(0)}(1+z)^{3}+\Omega_{\mathrm{DE}}^{(0)}\,\mathrm{exp}\left[3\int_{0}^{z}\frac{1+w(z')}{1+z'}\mathrm{d}z' \right]},
\end{equation}
where $\Omega_{i}^{(0)}$ is the present fraction of the critical
density in the form of component $i$, which can be radiation (r),
matter (m), baryonic matter (b), cold dark matter (cdm), curvature
(k) and dark energy (DE). The dark energy parameter $w(z)=w_{0}$
is a constant in the $\Lambda$CDM model. The mass density $\Omega_{m}(z)$
at a generic redshift is given by
\begin{equation}
\Omega_{m}(z)=\frac{\Omega_{m}^{(0)}(1+z)^{3}}{(H(z)/H_{0})^{2}}
\end{equation}
and $\Omega_{DE}^{(0)}=1-\Omega_{m}^{(0)}$. The angular diameter
distance $D_{A}(z)$ expressed in units of Mpc is
\begin{equation}
D_{A}(z)=\frac{c}{1+z}\int_{0}^{z}\frac{\mathrm{d}z'}{H(z')}
\end{equation}
and the growth function $G(k,z)$ is
\begin{equation}
G(k,z)=\frac{\delta_{m}(k,z)}{\delta_{m}(k,0)},
\end{equation}
where $\delta_{m}(k,z)$ is the matter density at a certain redshift
and scale. The growth rate is $f(k,z) \equiv\dot{\delta}_{m}(k,z)/(H(z)\delta_{m}(k,z))$
and in the $\Lambda$CDM model it can be approximated as scale independent,
$f(z)=\Omega_{m}^{\gamma}(z)$ \citep{peebles76,Lahav-etal:1991,polarski08,Linder:2005in}
where $\gamma\simeq0.55$ \citep{Peebles80,wang1998cluster}.

Given the set of cosmological parameters $\boldsymbol{\theta}=\{h,n_{s},\Omega_{\mathrm{b}}^{(0)},\Omega_{\mathrm{cdm}}^{(0)},w_{0}\}$,
the linear matter power spectrum at a redshift $z$ can then be written
as \citep{Amendola2010}
\begin{equation}
P(k,z;\boldsymbol{\theta})=A_{s}\,k^{q(n_{s})}\,T^{2}(k;\boldsymbol{\theta})\,G^{2}(k,z;\boldsymbol{\theta}),\label{eq:lin_spectrum}
\end{equation}
where $A_{s}$ is the amplitude of the spectrum, $q(n_{s})$ is a
function of the spectral index $n_{s}$, and $T(k;\boldsymbol{\theta})$ is the
transfer function that in a linear regime and in the $\Lambda$CDM model does not depend on redshift. Note that from now on, we highlight the dependence of the quantities $H(z; \boldsymbol{\theta})$, $D_A(z;\boldsymbol{\theta})$, $G^2(k,z;\boldsymbol{\theta})$ and $f(k,z;\boldsymbol{\theta})$ on the set $\boldsymbol{\theta}$ of cosmological parameters.

Since we observe the galaxy distribution in redshift space and not
directly the matter density in real space, we have to take into account
the bias and the redshift distortion. If we define the linear bias
as $\delta_{k,\mathrm{matter}}\cdot b(z,k)\equiv\delta_{k,\mathrm{galaxies}}$
and normalize the power spectrum to $\sigma_{8}$, then the following
equation follows
\begin{equation}
\delta_{\mathrm{obs}}(k,\mu)=\sigma_{8}b(k,z)\left[1+\frac{f(k,z;\boldsymbol{\theta})}{b(k,z;\boldsymbol{\theta})}\mu^2 \right]\delta_{k},\label{eq:redshift-distortions}
\end{equation}
where $\mu$ is the cosine of the angle between $\boldsymbol{k}$
and the line of sight. Although in this form the redshift distortion
is only valid in the flat-sky approximation, we apply it to a large-sky
survey as Euclid, following most of previous forecast works. The quantities
$G(k,z; \boldsymbol{\theta})$, $b(k,z)$, $f(k,z; \boldsymbol{\theta})$ are in general scale-dependent. Note
that the redshift distortions lead to an anisotropic observed power
spectrum. Finally, we can consider a scale independent residual shot-noise
term $P_{\mathrm{s}}(z)$ on top of the shot noise $1/n_{\mathrm{gal}}(z)$,
where $n_{\mathrm{gal}}(z)$ is the number density of galaxies.  This term is added in case of incomplete removal of the shot noise arising from the discrete sampling of galaxies. Thus,
the final expression is \citep{seo03}
\begin{equation}
P_{\mathrm{obs}}(k,\mu, z;\boldsymbol{\theta})=P_{\mathrm{s}}(z)\,+\,\sigma_{8}^{2}b^{2}(k,z)\left[1+\beta(k,z;\boldsymbol{\theta})\mu^{2}\right]^{2}P(k,z;\boldsymbol{\theta}),\label{eq:obs_spectrum_lin}
\end{equation}
where we defined $\beta(k,z;\boldsymbol{\theta})\equiv f(k,z;\boldsymbol{\theta})/b(k,z)$. We will consider $b$ to be scale-independent and, since the bias function
is undefined along with the parameters above, we add a bias parameter
for each bin and marginalize over it (see \cite{seo03} and Sec. \ref{sec:Conclusions})  assuming a  uniform prior.

If the set of parameters $\boldsymbol{\theta}$ represents our fiducial cosmological
model, varying it to a new set $\boldsymbol{\theta}'$ not only changes the shape
and amplitude of the spectrum, because the transfer and growth functions
$T(k; \boldsymbol{\theta})$, $G(k,z; \boldsymbol{\theta})$ are different, but new distortions are introduced
in the vector $\boldsymbol{k}$ and the volume in which the spectrum
is computed. By taking into account these modifications, called Alcock-Paczynski
(AP) effect, the observed spectrum can be formalized as \citep{Alcock:1979mp,seo03,Amendola2010}
\begin{equation}
P_{\mathrm{obs}}(k,\mu_{\mathrm{}},z;\boldsymbol{\theta})=P_{\mathrm{s}}(z)\,+\,\frac{H(z;\boldsymbol{\theta}')\cdot D_{\mathrm{A}}^{2}(z;\boldsymbol{\theta})}{H_{\mathrm{}}(z;\boldsymbol{\theta})\cdot D_{\mathrm{A}}^{2}(z;\boldsymbol{\theta}')}\sigma_{8}^{2}b^{2}(k,z)\left[1+\beta(k,z;\boldsymbol{\theta}')\mu_{\boldsymbol{\theta}'}^{2}\right]^{2}P(k_{\boldsymbol{\theta}'},z;\boldsymbol{\theta}'),\label{eq:obs_spectrum_lin_AP}.
\end{equation}
The transformations of $k=|\mathbf{k}|$ and $\mu$ are given
by the following equations:
\begin{equation}
k_{\boldsymbol{\theta}'}=\Upsilon\,k,\label{eq:AP_1}
\end{equation}
\begin{equation}
\mu_{\boldsymbol{\theta}'}=\frac{H(z; \boldsymbol{\theta}')\mu}{H(z; \boldsymbol{\theta})\Upsilon},\label{eq:AP_2}
\end{equation}
where
\begin{equation}
\Upsilon=\frac{\sqrt{H^{2}(z; \boldsymbol{\theta}')D_{\mathrm{A}}^{2}(z;\boldsymbol{\theta}')\mu^{2}-H^{2}(z,\boldsymbol{\theta})D_{\mathrm{A}}^{2}(z; \boldsymbol{\theta})[\mu^{2}-1]}}{H(z; \boldsymbol{\theta})D_{\mathrm{A}}(z; \boldsymbol{\theta}')}.\label{eq:AP_3}
\end{equation}
Now that we have \textit{\emph{introduced}} the observed matter power
spectrum, we will analyze how it is affected by the introduction of
a window function and the bin cross-correlations.
\section{Window function effect, correlations between bins, and bin redshift
uncertainty} \label{sec:Correlations-among-bins}

\subsection{Fisher information matrix for correlated bins} \label{sub:Fisher-information-matrix}

In the Fisher information matrix approach \citep{fisher1935logic,tegmark1997karhunen},
the Fourier coefficients $\delta_{\mathbf{k}}$ of the density contrast
are random variables, which are expected to have a Gaussian distribution
in the standard model of inflation and as long as they remain linear
\citep{lyth2009primordial}. If we call $\boldsymbol{\theta}$ the set of cosmological
parameters, then the likelihood is $P(\delta_{\mathbf{k}}|\boldsymbol{\theta})$.
By applying Bayes' theorem with an uniform prior we obtain the posterior
probability $\mathcal{L}(\boldsymbol{\theta}|\delta_{\mathbf{k}})\varpropto P(\delta_{\mathbf{k}}|\boldsymbol{\theta})$,
now viewed as a function of the parameters (for a detailed review
of this and what follows, see \citep{heavens2009statistical,trotta2008bayes,dodelson2003modern,Amendola2010}).
The probability $\mathcal{L}(\boldsymbol{\theta}|x)$ is also commonly called likelihood,
so we will also follow this notation. The Fisher matrix method approximates
such likelihood $\mathcal{L}(\boldsymbol{\theta}|x)$ around its peak by a Gaussian
correlated distribution of parameters $\boldsymbol{\theta}$. The Fisher matrix
is defined as the inverse of the parameter covariance matrix of the
distribution and it encodes then the Gaussian uncertainties $\sigma_{\boldsymbol{\theta}}$
on the parameters. The maximum of $\mathcal{L}(\boldsymbol{\theta}|x)$ can be
found using efficient numerical algorithms, e.g. Newton-Raphson \citep{press2007numerical},
and the symmetric Fisher matrix can be obtained by sampling $\mathcal{L}(\boldsymbol{\theta}|\delta_{\mathbf{k}})$
$2N_{\boldsymbol{\theta}}$ times, where $N_{\boldsymbol{\theta}}$ is the number of parameters.
Here we will use the Fisher matrix tools to propagate the uncertainties
on the cosmological parameters when data are not available yet, i.e.
when we are \textit{\emph{forecasting a future galaxy redshift survey}}.
This is possible because in a forecast we can fix the maximum of $\mathcal{L}(\boldsymbol{\theta}|\delta_{\mathbf{k}})$
at the fiducial cosmological parameters $\boldsymbol{\theta}_{0}$ given by previous
data.

The standard Fisher matrix approach typically employed in forecast
studies for galaxy clustering is the following \citep{seo03}. After
dividing the survey space in $N$ redshift bins along the line of
sight, a set of $N$ coefficients $\boldsymbol{\delta}_{\mathbf{k}}^{\mathrm{}}=\{\delta_{\mathbf{k}}^{(1)},...\,,\delta_{\mathbf{k}}^{(N)}\}$
is deduced from the data. All the redshift dependent quantities, e.g.
$G(z;\boldsymbol{\theta}),b(z),f(z;\boldsymbol{\theta})$, are taken at the median bin redshift (from now
on, we ignore their possible $k$-dependence). The random variables
assigned to the coefficients are statistically independent, because
bin cross-correlation spectra $V\left\langle \delta_{\mathbf{k}}^{(i)}\delta_{\mathbf{-k}}^{(j)}\right\rangle $
for $i\neq j$ are ignored and set to zero. Also, the effects of the
window function are neglected, thus the observed power spectrum in
Eq. (\ref{eq:obs_spectrum_lin_AP}) for each bin $i$ is $\mbox{V\ensuremath{\left\langle \delta_{\mathbf{k}}^{(i)}\delta_{\mathbf{-k}}^{(i)}\right\rangle \stackrel{!}{=}}P(k,\ensuremath{\mu},\ensuremath{z_{i}})}$.
The definition of the power spectrum in Eq. (\ref{eq:def_pow_spectr})
shows that modes at different wave vectors $\boldsymbol{k}$ are independent
if we assume an homogeneous field, i.e. $\left\langle \delta_{\mathbf{k}}\delta_{\mathbf{-k'}}\right\rangle =0$
for $\boldsymbol{k\neq k'}$. The likelihood is then a multivariate
Gaussian distribution in the random variables given by the product
of the Gaussian distributions of the $N$ bins: \citep{Amendola2010}
\begin{equation}
\mathcal{L}(\boldsymbol{\theta})=\prod_{i=1}^{N}\prod_{l}^{M}\left\{ \frac{1}{\sqrt{2\pi} \ensuremath{\Delta_{\mathbf{k}_{l},i}}}\mathrm{exp}\left[-\frac{1}{2}\frac{\left|\delta_{\mathbf{k}_{l}}^{(i)}\right|^{2}}{\Delta_{\mathbf{k}_{l},i}^{2}}\right]\right\} ,\label{eq:uncorr_likelihood}
\end{equation}
where for simplicity we have only a set of $M$ discrete wave vectors
$\mathbf{k}_{l}$ and where the variance is
\begin{equation}
\mbox{\ensuremath{\Delta_{\mathbf{k}_{l},i}^{2}}=\ensuremath{V_{i}\left\langle \delta_{\mathbf{k}_{l}}^{(i)}\delta_{\mathbf{-\mathbf{k}}_{l}}^{(i)}\right\rangle }+\ensuremath{\frac{1}{n_{\mathrm{gal}}(z_i)}}}.
\end{equation}
The second term in the variance is the shot noise \citep{seo03} and
the coefficients $n_{\mathrm{gal}}(z_i)$ are the densities of galaxies
in the $i$-th redshift bin, assumed to be constant inside a bin.

The resulting element of the standard Fisher matrix for two parameters
$\theta_{\alpha}$ and $\theta_{\beta}$ is then: \citep{seo03,tegmark1997measuring}
\begin{equation}
F_{\alpha\beta}=\sum_{i=1}^{N}\frac{1}{8\pi^{2}}\int_{-1}^{1}d\mu\int_{k_{\mathrm{min}}}^{k_{\mathrm{max}}}k^{2}dk\left(\frac{\partial\mathrm{ln}P_{\mathrm{obs}}(k,\mu,z_{i};\boldsymbol{\theta})}{\partial\theta_{\alpha}}\frac{\partial\mathrm{ln}P_{\mathrm{obs}}(k,\mu,z_{i};\boldsymbol{\theta})}{\partial\theta_{\beta}}\right)V_{\mathrm{eff}}^{i}(k,\mu;\boldsymbol{\theta}),\label{eq:uFM_integral}
\end{equation}
where $P_{\mathrm{obs}}$ was given in \eqref{eq:obs_spectrum_lin_AP} and
\begin{equation}
V_{\mathrm{eff}}^{i}(k,\mu;\boldsymbol{\theta})=\left[\frac{n_{i}P_{\mathrm{obs}}(k,\mu,z_{i};\boldsymbol{\theta})}{n_{i}P_{\mathrm{obs}}(k,\mu,z_{i};\boldsymbol{\theta})+1}\right]^{2}f_{\mathrm{sky}}V_{i}.\label{eq:eff_volume}
\end{equation}

If we want to consider the correlation function given by pairs of
galaxies belonging to different redshift bins, the cross-correlation
spectra $\sqrt{V_{i}V_{j}}\left\langle \delta_{\mathbf{k}}^{(i)}\delta_{\mathbf{-k}}^{(j)}\right\rangle $
for $i\neq j$ need also to be considered and the coefficients $\delta_{\mathbf{k}}^{(i)\mathrm{}}$
are no longer statistically independent. The cross-correlation spectra
represent indeed the off-diagonal elements of the covariance matrix
(see Sec. \ref{sub:Windowed-and-cross-correlation} for more details):
\begin{equation}
\mbox{\ensuremath{C_{ij}}(\ensuremath{\mathbf{k}})=\ensuremath{\sqrt{V_{i}V_{j}}\left\langle \delta_{\mathbf{k}}^{(i)}\delta_{-\mathbf{k}}^{(j)}\right\rangle }+\ensuremath{\frac{\delta_{ij}}{n_{\mathrm{gal}}(z_i)}}}.\label{eq:covariance_matrix}
\end{equation}
Equation (\ref{eq:uncorr_likelihood}) is then rewritten in the form
of a Gaussian correlated distribution for the bins:
\begin{equation}
\mathcal{L}(\boldsymbol{\theta})=\prod_{l}^{M}\left\{ \frac{1}{(2\pi)^{N/2}\sqrt{\det\mathbf{C}(\mathbf{k}_l)}}\exp\left[-\frac{1}{2}\sum_{i,j}^{N}\delta_{\mathbf{k}_{l}}^{(i)}\,C_{ij}^{-1}(\mathbf{k}_{l})\,\delta_{-\mathbf{k}_{l}}^{(i)}\right]\right\} ,\label{eq:correlated_likelihood}
\end{equation}
where bold and capital letters represent $N\times N$ matrices over
the bins and we defined the following notation:
\begin{equation}
P_{\mathrm{corr}}^{i,j}(\mathbf{k}_{l})\equiv\sqrt{V_{i}V_{j}}\left\langle \delta_{\mathbf{k}_{l}}^{(i)}\delta_{\mathbf{-k}_{l}}^{(j)}\right\rangle.
\end{equation}
 An important remark is that modes at different wave vectors $\mathbf{k}_{l}$
are still assumed independent in Eq. (\ref{eq:correlated_likelihood}),
but the introduction of the window function effects breaks the homogeneity
of the density field, thus this assumption is no longer true. We will
discuss again this point in Sec. \ref{sub:Windowed-and-cross-correlation}.

The Fisher matrix element in the correlated case becomes then

\begin{align}
F_{\alpha\beta} & =\frac{1}{2}\sum_{i,j,h,g}^{N}\sum_{l}^{M}\frac{\partial P_{\mathrm{corr}}^{i,j}(\mathbf{k}_{l};\boldsymbol{\theta})}{\partial\theta_{\alpha}}\,C_{jh}^{-1}(\mathbf{k}_{l})\,\frac{\partial P{}_{\mathrm{corr}}^{h,g}(\mathbf{k}_{l};\boldsymbol{\theta})}{\partial\theta_{\beta}}\,C_{gi}^{-1}(\mathbf{k}_{l})f_{\mathrm{sky}}\sqrt[4]{V_{\mathrm{}}^{i}V_{\mathrm{}}^{j}V_{\mathrm{}}^{h}V_{\mathrm{}}^{g}}\label{eq:cFM}\\
 & =\frac{1}{8\pi^{2}}\int_{k_{\mathrm{min}}}^{k_{\mathrm{max}}}\mathrm{d}k\,k^{2}\int_{-1}^{+1}\mathrm{d}\mu\,\,f_{\mathrm{sky}}\mathrm{Tr}\left[\frac{\partial\mathbf{P}_{\mathrm{corr}}(k,\mu;\boldsymbol{\theta})}{\partial\theta_{\alpha}}\times\hat{\mathbf{C}}^{-1}(k,\mu)\times\frac{\partial\mathbf{P}_{\mathrm{corr}}(k,\mu;\boldsymbol{\theta})}{\partial\theta_{\beta}}\times\mathbf{\hat{C}}^{-1}(k,\mu)\right],\label{eq:cFM_integral}
\end{align}
where in the last passage for convenience we included the survey comoving
volumes in the inverse covariance matrix: $\mbox{\ensuremath{\hat{C}_{ij}^{-1}}(k,\ensuremath{\mu})\ensuremath{\equiv C_{ij}^{-1}}(k,\ensuremath{\mu})\ensuremath{\sqrt[4]{V_{i}V_{j}}}}$.

\subsection{Redshift as an additional parameter} \label{sub:Redshift-as-an}

In previous Sec. \ref{sub:Fisher-information-matrix}, we assumed
the redshift-dependent functions $G(z;\boldsymbol{\theta}),b(z),f(z;\boldsymbol{\theta}),V(z)$ to be constant
inside bins and computed them at the median bin redshift. When the
bins are wide in redshift, for instance with $\Delta z\thickapprox0.5$,
then this assumption is not a good approximation. To test the impact
of this simplification, let us now assign a random variable to the
value of the redshift at which we evaluate the redshift-dependent
functions. Then the likelihood in the uncorrelated case becomes
\begin{equation}
\mathcal{L}(\boldsymbol{\theta})=\prod_{i=1}^{N}\left\{ \prod_{l}^{M}\left\{ \frac{1}{\sqrt{2\pi} \ensuremath{\Delta_{\mathbf{k}_{l},i}} } \mathrm{exp}\left[-\frac{1}{2}\frac{\left|\delta_{\mathbf{k}_{l}}^{(i)}\right|^{2}}{\Delta_{\mathbf{k}_{l},i}^{2}}\right]\right\} \frac{1}{\sqrt{2\pi}\sigma_{i}}\exp\left[-\frac{1}{2}\frac{(z_{i}-\hat{z}_{i})^{2}}{\sigma_{i}^{2}}\right]\right\} ,\label{eq:uncorr_likelihood_redshift}
\end{equation}
where for each bin, $z_{i}$ are parameters which have a Gaussian
distribution centered at the median bin redshift $\hat{z}_{i}$ and
a standard deviation $\sigma_{i}=\Delta z/6$ equal to one-sixth of
the bin redshift width. With this choice, the values of parameters
$z_{i}$ have 99.7\% probability to lie within the $i$-th bin, i.e.
within the three standard deviations range of the probability distribution.
This Gaussian distribution can be seen as a prior on the new parameters
$z_{i}$.

The first part of the likelihood in Eq. (\ref{eq:uncorr_likelihood_redshift})
depends on the cosmological parameters and on $z_{i}$, whereas the
second part depends on the $z_{i}$ only. The logarithm of the likelihood
is:
\begin{equation}
-\mathrm{ln}\mathcal{L}=\frac{\mathrm{ln}(2\pi)}{2}\left(MN+N\right)+\sum_{i}\left\{ \sum_{l}\mathrm{ln}\Delta_{\mathbf{k}_{l},i}+\sum_{l}\frac{1}{2}\left(\frac{\left|\delta_{\mathbf{k}_{l}}^{(i)}\right|^2}{\Delta_{\mathbf{k}_{l},i}^{2}}\right)^{2}+\mathrm{ln}\sigma_{i}+\frac{1}{2}\left(\frac{z_{i}-\hat{z}_{i}}{\sigma_{i}}\right)^{2}\right\} .
\end{equation}
The Fisher matrix $\hat{F}_{AB}$ is then an extended version of the
old Fisher matrix $F_{\alpha\beta}$ in Eq. (\ref{eq:uFM_integral}),
such that $A,B$ are indices ranging from 1 to $N_{\mathrm{cosmo}}+2N$,
where $N$ is the number of bins and $N_{\mathrm{cosmo}}$ is the
number of cosmological parameters.

For $A,B$ ranging from 1 to $N_{\mathrm{cosmo}}+N$ (i.e., the cosmological
parameters plus one value of bias for each bin), the Fisher matrix
$\hat{F}_{AB}$ is equal to $F_{AB}$. For $A,B$ ranging over $[N_{\mathrm{cosmo}}+N+1,2N]$,
the Fisher matrix elements are
\begin{equation}
\hat{F}{}_{AB}=\frac{\delta_{AB}}{\sigma_{\hat{A}}^{2}}+\sum_{i=1}^{N}\frac{1}{8\pi^{2}}\int_{k_{\mathrm{min}}}^{k_{\mathrm{max}}}\mathrm{d}k\,k^{2}\int_{-1}^{+1}\mathrm{d}\mu\left(\frac{\partial\mathrm{ln}P_{\mathrm{obs}}(k,\mu,z_{i};\boldsymbol{\theta})}{\partial z_{\hat{A}}}\frac{\partial\mathrm{ln}P_{\mathrm{obs}}(k,\mu,z_{i};\boldsymbol{\theta})}{\partial z_{\hat{B}}}\right)V_{\mathrm{eff}}^{i}(k,\mu),
\end{equation}
where $\hat{A}=A-N_{\mathrm{cosmo}}-N-1$ and $\hat{B}=B-N_{\mathrm{cosmo}}-N-1$.
For $A$ ranging over $[N_{\mathrm{cosmo}}+N+1,2N]$ and $B$ ranging
from 1 to $N_{\mathrm{cosmo}}+N$, the mixed terms of the Fisher matrix
are
\begin{equation}
\hat{F}{}_{AB}=\sum_{i=1}^{N}\frac{1}{8\pi^{2}}\int_{k_{\mathrm{min}}}^{k_{\mathrm{max}}}\mathrm{d}k\,k^{2}\int_{-1}^{+1}\mathrm{d}\mu\left(\frac{\partial\mathrm{ln}P_{\mathrm{obs}}(k,\mu,z_{i};\boldsymbol{\theta})}{\partial z_{\hat{A}}}\frac{\partial\mathrm{ln}P_{\mathrm{obs}}(k,\mu,z_{i};\boldsymbol{\theta})}{\partial\boldsymbol{\theta}_{B}}\right)V_{\mathrm{eff}}^{i}(k,\mu).
\end{equation}
Similar calculations lead to corresponding results for the correlated
Fisher approach described in Eq. (\ref{eq:cFM}). In Sec. \ref{sub:Assumption-effects-with}
we will show that if we introduce this bin redshift uncertainty, the
values of the uncertainties on the cosmological parameters are larger
for wider bins in redshift.

\subsection{Window function and bin cross-correlation spectra} \label{sub:Windowed-and-cross-correlation}

We now describe the cross-correlation spectrum $\mbox{\ensuremath{\mathbf{P}_{\mathrm{corr}}^{i,j}}(\ensuremath{\mathbf{k}_{l}})\ensuremath{\equiv\sqrt{V_{i}V_{j}}\left\langle \delta_{\mathbf{k}_{l}}^{(i)}\delta_{\mathbf{-k}_{l}}^{(j)}\right\rangle }}$
introduced previously and the recovery of the standard power spectrum
in the limit of an \textit{\emph{infinite survey}}. Furthermore, we
justify the shot noise term introduced in the covariance matrix, Eq.
(\ref{eq:covariance_matrix}). With the introduction of the top-hat
survey-bin window function $W_{i}(\mathbf{x})$, such that $\int W_{i}(\mathbf{x})d^{3}x=V_{i}$,
the coefficients $\delta_{\mathbf{k}}^{(i)}$ can be expanded as:
\begin{align}
\delta_{\mathbf{k}}^{(i)} & =\frac{1}{V_{i}}\int\delta(\mathbf{x})W_{i}(\mathbf{x})e^{-i\mathbf{k}\cdot\mathbf{x}}d^{3}x\\
 & =\frac{1}{V_{i}}\int\left(\frac{\rho(\mathbf{x})}{\rho_{0}}\right)W_{i}(\mathbf{x})e^{-i\mathbf{k\cdot}\mathbf{x}}d^{3}x-\tilde{W_{i}}(\mathbf{k}),\label{eq:contrast_coeff}
\end{align}
where
\begin{align}
\tilde{W}_{i}(\mathbf{k}) & =\frac{1}{V_{i}}\int W_{i}(\mathbf{x})e^{i\mathbf{k}\cdot\mathbf{x}}d^{3}x.\label{eq:def_fourier_window-1}
\end{align}
It follows that the cross-correlations between the Fourier coefficients of the density field can be written as \citep{FKP}:
\begin{align}
\sqrt{V_{i}V_{j}}\left\langle \delta_{\mathbf{k}}^{(i)}\delta_{-\mathbf{k'}}^{(j)}\right\rangle =\frac{\sqrt{V_{i}V_{j}}}{(2\pi)^{3}}\int P(k'')\tilde{W}_{i}(\mathbf{k}-\mathbf{k}'')\tilde{W}_{j}(\mathbf{k}'-\mathbf{k}'')d^{3}k''.\label{eq:corr_spectra_ALL}
\end{align}
\begin{center}
\begin{figure}
\centering
\includegraphics[scale=0.72]{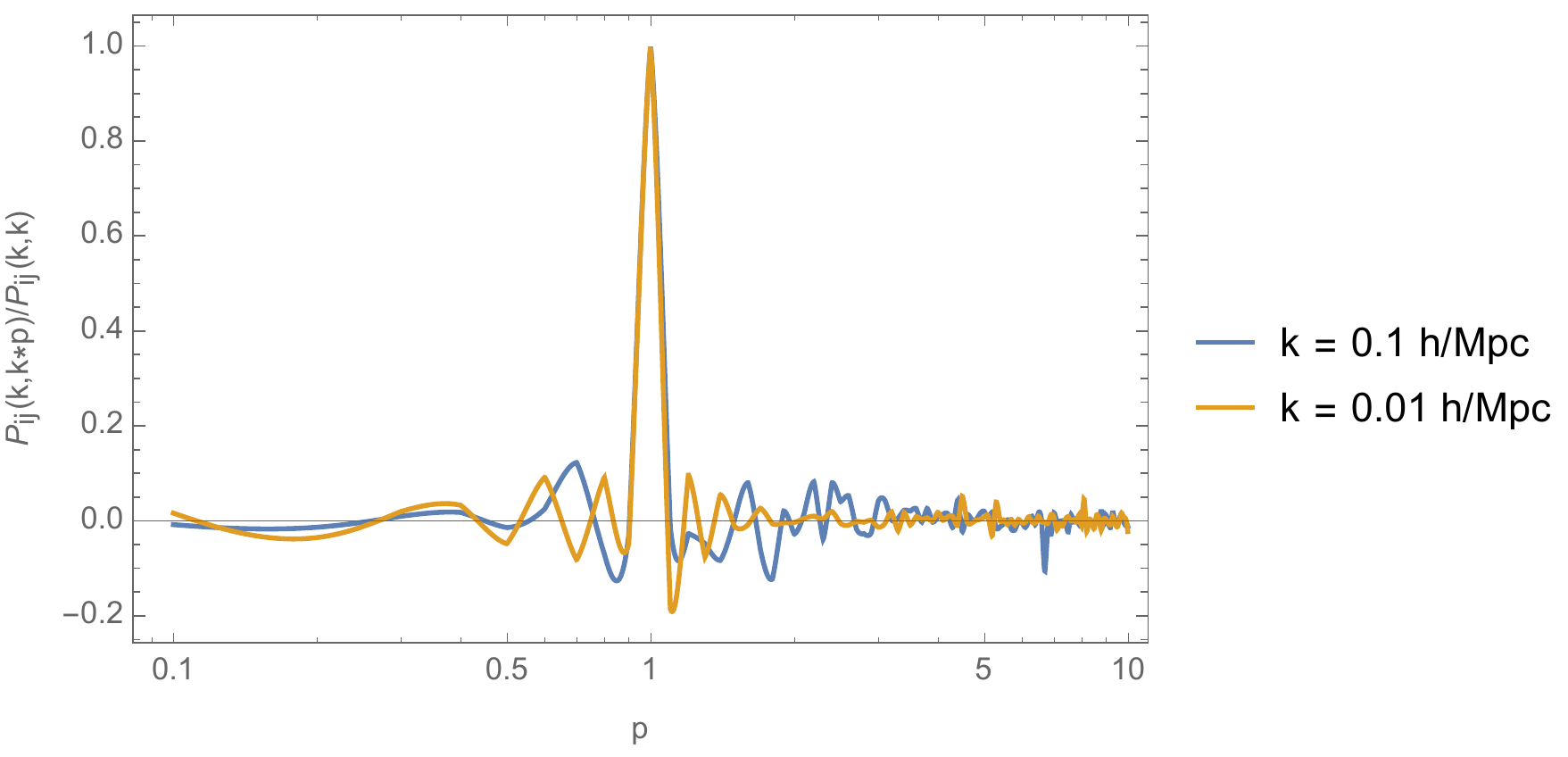}\caption{An example of the ratio between expression \eqref{eq:corr_spectra_ALL} for $\boldsymbol{k\neq k'}$ and the same expression for $\boldsymbol{k = k'}$. Specifically, the plot shows the ratio between the power spectrum $P_{ij}(\bold{k}, \bold{k'})=\sqrt{V_{i}V_{j}}\left\langle \delta_{\mathbf{k}}^{(i)}\delta_{-\mathbf{k'}}^{(j)}\right\rangle $ calculated between two contiguous redshift bins (one ranging from $z=1.05$ to $z=1.15$ and the other from $z=1.15$ to $z=1.25$, corresponding to two typical central bins for \textit{Euclid}), at values $k = 0.1 h/$Mpc or $k = 0.01 h/$Mpc and $k'= p k$, with $p$ a dimensionless constant ranging from 0.1 to 10. For both reference values of $k$ the correlations peak in correspondence of $p=1$, showing that correlations between different wave vectors are indeed much smaller than those between the same wave vector. The power spectrum used for the calculation has the typical form $\frac{k}{1+k^4}$ and $\bold{k'}$ is considered to be parallel to $\bold{k}$ for simplicity, to perform easily the calculations and give a taste of the different magnitudes of the correlations.}
\label{fig:kpk}
\end{figure}
\end{center}
This equation shows that this mode-mode correlation (i.e. $\boldsymbol{k\neq k'}$) does not vanish both for $i=j$ and $i\neq j$. Nevertheless, from now on we will only consider correlations between the same wave vectors, as the ones between different $\boldsymbol{k}$ values lead to more difficult numerical computations and need a new formalization in the Fisher matrix approach. They are in any case smaller than the correlations between the same wave vectors as illustrated in Fig.~\ref{fig:kpk}, which justifies our choice.

\noindent Correlations between the same wave vectors take the form of a convolution of $P(k)$:
\begin{equation}
\sqrt{V_{i}V_{j}}\left\langle \delta_{\mathbf{k}}^{(i)}\delta_{\mathbf{-k}}^{(j)}\right\rangle =\frac{\sqrt{V_{i}V_{j}}}{(2\pi)^{3}}\int P(k')\tilde{W}_{i}(\mathbf{k}-\mathbf{k}')\tilde{W}_{j}(\mathbf{k}-\mathbf{k}')d^{3}k'\equiv\frac{\sqrt{V_{i}V_{j}}}{(2\pi)^{3}}\int d^{3}k'P(k')\tilde{Q}_{ij}(\mathbf{k-k'}),\label{eq:corr_spectra}
\end{equation}
where $\tilde{Q}_{ij}(\mathbf{k})\equiv\tilde{W}_{i}(\mathbf{k})\tilde{W}_{j}(\mathbf{k})$.
This convolution becomes radial when we restrict ourselves to a top-hat
spherical window function describing a redshift bin between the comoving
distances $R_{i}$ and $R_{i}+\Delta_{i}$. In fact, the Fourier transform
of this specific window function is:
\begin{align}
\tilde{W}_{i}(|\mathbf{k-k'}|,R_{i},\Delta_{i}) & =\frac{1}{V_{i}}\int W_{i}(|\mathbf{x}|)e^{i(\mathbf{k-k'})\cdot\mathbf{x}}d^{3}x\label{eq:def_fourier_window}\\
 & =\frac{4\pi}{V_{i}}\int_{R_{i}}^{R_{i}+\Delta_{i}}\frac{\sin\left(|\mathbf{k-k'}|x\right)}{|\mathbf{k-k'}|x}\,x^{2}dx\nonumber \\
 & =\frac{4\pi}{V_{i}}\left[H\left(|\mathbf{k-k'}|(R_{i}+\Delta_{i})\right)-H\left(|\mathbf{k-k'}|R_{i}\right)\right],\label{eq:fourier_window_anal}
\end{align}
where
\begin{equation}
H(kr)=\frac{\sin kr-kr\cos kr}{k^{3}}.
\end{equation}
This shows that the quantity $\tilde{Q}_{ij}(\mathbf{k})$ defined
in Eq. (\ref{eq:corr_spectra}) will depend only on the modulus of
$\mathbf{k}$, thus the convolution becomes radial and the cross-correlation
spectra will depend only on the modulus $k$.

In the limit of an infinite survey one recovers the standard power
spectrum and\emph{ }\textit{\emph{vanishing}} cross-correlations.
In this limit the window function in Fourier space can indeed be approximated
as a Dirac delta, thus Eq. (\ref{eq:corr_spectra_ALL}) with $i=j$ gives
the standard (auto)-spectrum:
\begin{eqnarray}
V_{i}\left\langle \delta_{k}^{(i)}\delta_{-k}^{(i)}\right\rangle  & = & \frac{V_{i}}{(2\pi)^{3}V_{i}^{2}}\int P(k')e^{i(\mathbf{k}'-\mathbf{k})\cdot\mathbf{x}}e^{-i(\mathbf{k}-\mathbf{k'})\cdot\mathbf{y}}d^{3}k'd^{3}xd^{3}y\\
 & = & \frac{1}{V_{i}}\int P(k')\delta_{D}(\mathbf{k}'-\mathbf{k})e^{-i(\mathbf{k}-\mathbf{k'})\cdot\mathbf{y}}d^{3}k'd^{3}y\\
 & = & P(k).\label{eq:large_limit}
\end{eqnarray}
 Furthermore, in the limit of bins that are far away from each other,
the correlation function in Eq. (\ref{eq:corr_spectra_ALL}) for $i\neq j$
is computed at large distances $|\mathbf{x-y}|$ where its value is
close to zero. Thus, in this limit we find vanishing correlations $\sqrt{V_{i}V_{j}}\left\langle \delta_{\mathbf{k}}^{(i)}\delta_{-\mathbf{k'}}^{(j)}\right\rangle $ (see Eq. (\ref{eq:corr_spectra_ALL})).

Finally, we derive the shot-noise term introduced in the covariance
matrix in Eq. (\ref{eq:correlated_likelihood}). So far we analyzed
a continuum field, but for a galaxy survey we should rather consider
a discrete distribution of $N_{i}$ particles located at positions
$\mathbf{x}_{l}$ inside the $i$-th bin. From Eq. (\ref{eq:contrast_coeff})
we rewrite the Fourier coefficients as
\begin{align}
\delta_{\mathbf{k}}^{(i)} & =\frac{1}{V_{i}}\int\frac{\sum_{l}\delta_{D}(\mathbf{x}-\mathbf{x}_{l})}{\rho_{0}}W_{i}(\mathbf{x})e^{-i\mathbf{k\cdot}\mathbf{x}}d^{3}x-\tilde{W_{i}}(\mathbf{k})\nonumber \\
 & =\frac{1}{N_{i}}\sum_{l}W_{i}(\mathbf{x}_{l})e^{-i\mathbf{k\cdot}\mathbf{x}_{l}}-\tilde{W_{i}}(\mathbf{k})
\end{align}
and since $\langle\delta(k)\rangle=0$ we have
\begin{equation}
\frac{1}{N_{i}}\left\langle \sum_{l}W_{i}(\mathbf{x}_{l})e^{-i\mathbf{k\cdot}\mathbf{x}_{l}}\right\rangle =\tilde{W_{i}}(\mathbf{k}).
\end{equation}
The cross-correlations are then
\begin{align}
\sqrt{V_{i}V_{j}}\left\langle \delta_{k}^{(i)}\delta_{-k}^{(j)}\right\rangle & =\sqrt{V_{i}V_{j}}\left\langle \left(\frac{1}{N_{i}}\sum_{l}W_{i}(\mathbf{x}_{l})e^{-i\mathbf{k\cdot}\mathbf{x}_{l}}-\tilde{W_{i}}(\mathbf{k})\right)\left(\frac{1}{N_{j}}\sum_{m}W_{j}(\mathbf{x}_{m})e^{+i\mathbf{k\cdot}\mathbf{x}_{m}}-\tilde{W_{j}}(\mathbf{-k})\right)\right\rangle \\
 & =\frac{\delta_{ij}V_{i}}{N_{i}}+\frac{\sqrt{V_{i}V_{j}}}{N_{i}N_{j}}\left\langle \sum_{l\not=m}W_{i}(\mathbf{x}_{l})W_{j}(\mathbf{x}_{m})e^{-i\mathbf{k\cdot}(\mathbf{x}_{l}-\mathbf{x}_{m})}\right\rangle -\sqrt{V_{i}V_{j}}\tilde{W_{i}}(\mathbf{k})\tilde{W_{j}}(\mathbf{-k})\\
 & \equiv\delta_{ij}P_{N}+P_{\mathrm{filtered}}(k)-P_{W}(k).
\end{align}
This coincides with the previous result except for the shot noise
$P_{N}=V_{i}/N_{i}=1/n_{\mathrm{gal}}(z_i)$. As we can note, this
term is not present for cross-correlation spectra between different
bins.

\begin{table}
\centering{}%
\begin{tabular}{c|c|c|c}
Bin's redshift range & Average redshift & Density $n(z_{i})\cdot10^{-3}$ & Bias $b(z_{i})$\tabularnewline
\hline
\hline
0.65 - 0.75 & $z_{1}=0.7$ & 1.25 & 1.30\tabularnewline
0.75 - 0.85 & $z_{2}=0.8$ & 1.92 & 1.34\tabularnewline
0.85 - 0.95 & $z_{3}=0.9$ & 1.83 & 1.38\tabularnewline
0.95 - 1.05 & $z_{4}=1.0$ & 1.68 & 1.41\tabularnewline
1.05 - 1.15 & $z_{5}=1.1$ & 1.51 & 1.45\tabularnewline
1.15 - 1.25 & $z_{6}=1.2$ & 1.35 & 1.48\tabularnewline
1.25 - 1.35 & $z_{7}=1.3$ & 1.20 & 1.52\tabularnewline
1.35 - 1.45 & $z_{8}=1.4$ & 1.00 & 1.55\tabularnewline
1.45 - 1.55 & $z_{9}=1.5$ & 0.80 & 1.58\tabularnewline
1.55 - 1.65 & $z_{10}=1.6$ & 0.58 & 1.61\tabularnewline
1.65 - 1.75 & $z_{11}=1.7$ & 0.38 & 1.64\tabularnewline
1.75 - 1.85 & $z_{12}=1.8$ & 0.35 & 1.67\tabularnewline
1.85 - 1.95 & $z_{13}=1.9$ & 0.21 & 1.70\tabularnewline
 1.95 - 2.05 & $z_{14}=2.0$ &  0.11 &  1.73\tabularnewline
\end{tabular}\caption{Specifications of the survey Euclid: 14 bins from redshift $0.65$ to $2.05$ with depth $\Delta z=0.1$ \citep{1206.1225,euclidredbook}. For each bin, the table lists the redshift ranges, the average redshift, the galaxy density, and the fiducial values of the bias. The number density of galaxies $n(z)$ is estimated using the latest empirical data (see Table 3 in \citep{1206.1225} and Figure 3.2 in \citep{euclidredbook}). We follow the commonly employed procedure in galaxy clustering Fisher matrix forecasts, such that within each bin the galaxy density $n(z_i)$ is assumed to be constant (see \citep{seo03,2005MNRAS.357..429A,wang2006dark,albrecht2009findings,di2012simultaneous,wang2012robust,xu2014forecasts}).}
\label{table:survey_specs}
\end{table}

\begin{figure}
\begin{minipage}[t]{0.48\columnwidth}%
\begin{center}
\includegraphics[scale=0.42]{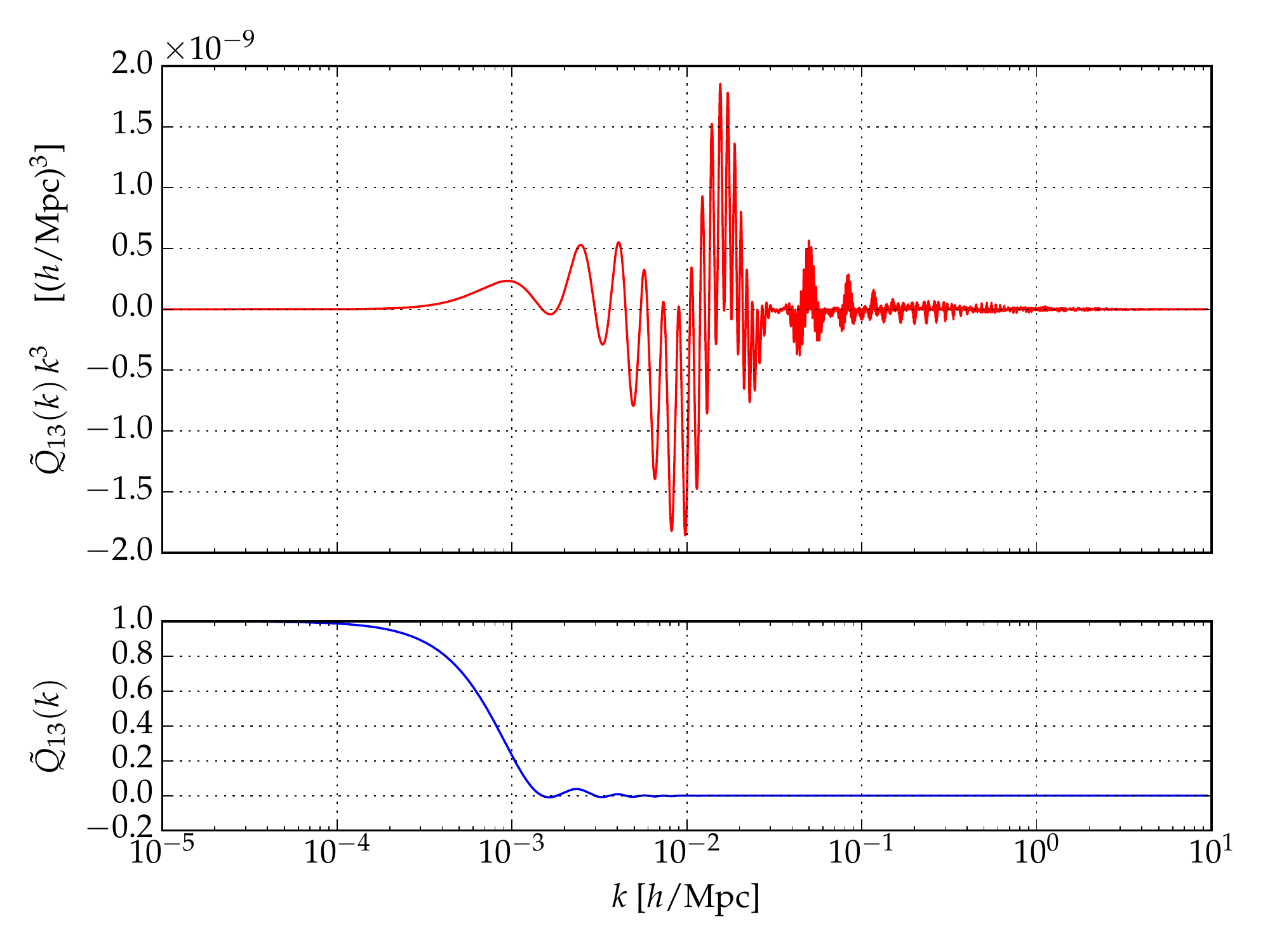}\caption{We consider the first and third bins of Euclid (Table \ref{table:survey_specs}).
We plot $\tilde{Q}_{13}(k)$ (blue plot) and $\tilde{Q}_{13}(k)k^{3}$
(red plot), as defined in Eq. (\ref{eq:corr_spectra}). The integrand
of the logarithmic version of the integral in Eq. (\ref{eq:angular_integral})
is proportional to $\tilde{Q}_{13}(k)k^{3}$ multiplied by the power
spectrum $P(k)$. The plot of this rapidly oscillating integrand gives
an idea of why it is hard to integrate it accurately.}
\label{fig:Q_ij plot}
\par\end{center}%
\end{minipage}\hfill{}%
\begin{minipage}[t]{0.48\columnwidth}%
\begin{center}
\includegraphics[scale=0.42]{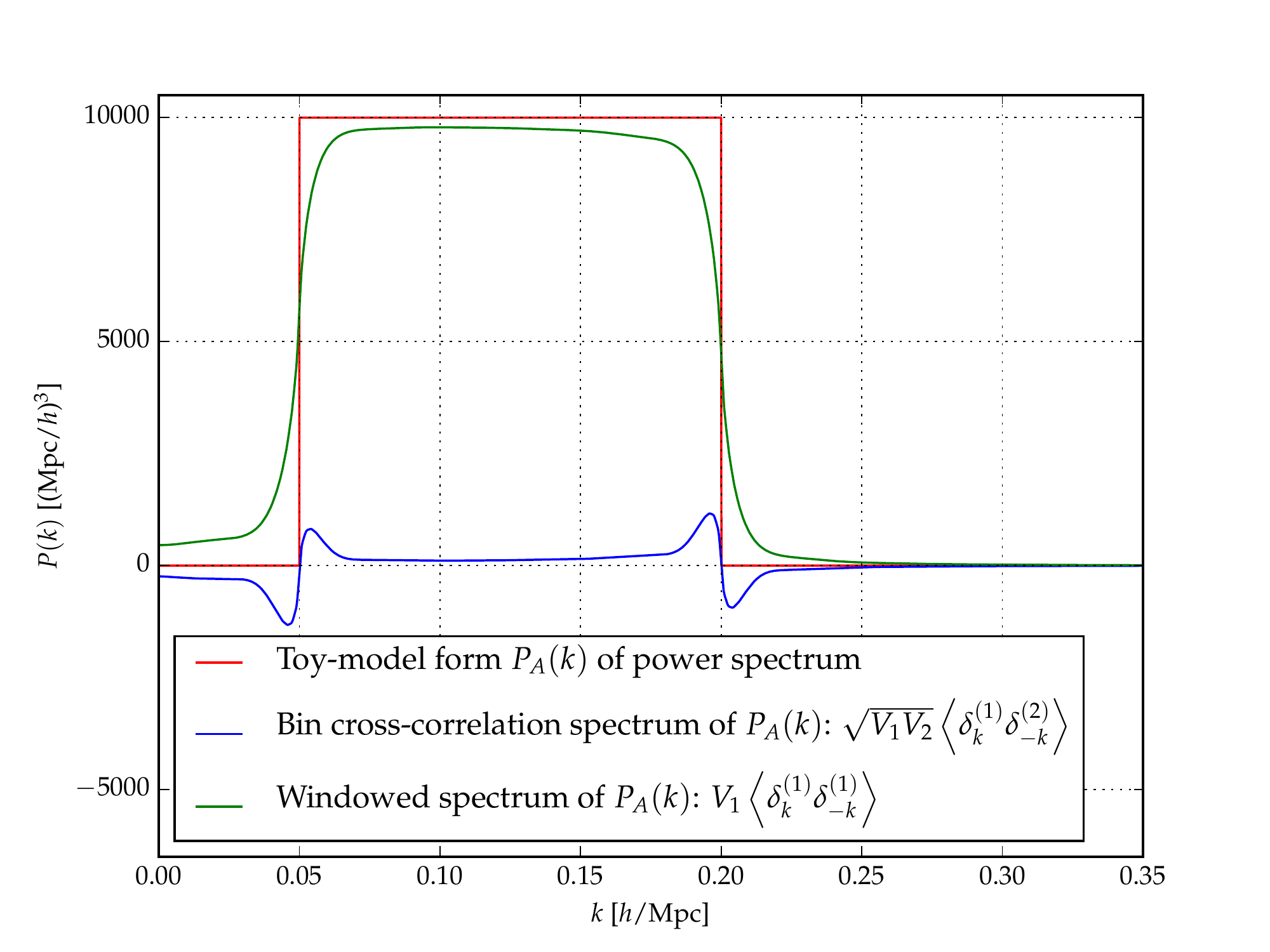}\caption{A toy-model form $P_{A}(k)$ (red plot) of the power spectrum is considered:
$P_{A}(k)=10^{4}$ (Mpc/$h$)$^{3}$ for $k\in[0.05,0.2]$ $h$/Mpc
and zero otherwise. We consider the first and second bins of Euclid
(Table \ref{table:survey_specs}). The green and the blue plots represent
$V_{1}\left\langle \delta_{k}^{(1)}\delta_{-k}^{(1)}\right\rangle $
and $\sqrt{V_{1}V_{2}}\left\langle \delta_{k}^{(1)}\delta_{-k}^{(2)}\right\rangle $,
respectively (see Eq. \ref{eq:angular_integral}). They are computed
using a radial FFT algorithm with parameters $\mathcal{N}=7000$,
$\mathcal{R}=1.5$ $h$/Mpc (App. \ref{sec:FFT-algorithm-for}). }
\label{fig:plot_analytical_spectrum}
\par\end{center}%
\end{minipage}
\end{figure}

\section{Numerical evaluation of theoretical observed windowed spectra and observed bin cross-correlation
spectra} \label{sec:computing-obs-spectr}

In this section, we analyze how the redshift distortions modify the
cross-correlation spectra and we tune the parameters of the radial
FFT algorithm based on three toy-model forms of power spectra for
which the convolution integral can be solved analytically. Assuming
a linear power spectrum, we now introduce the redshift distortions
given by Eq. (\ref{eq:redshift-distortions}) in the cross-correlation
spectra, similarly to what we did in Sec. \ref{sec:The-observed-matter}.
If we assume that quantities $G(z;\boldsymbol{\theta}),b(z),f(z;\boldsymbol{\theta})$ are scale independent
as we did previously, Eq. (\ref{eq:obs_spectrum_lin_AP}) can then
be generalized as
\begin{eqnarray}
P_{\mathrm{obs}}^{\mathrm{}}(k,\mu,z_{i},z_{j};\boldsymbol{\theta}) & = & \sigma_{8}^{2}b(z_i) b(z_j) \sqrt{\left[\frac{H(z_{i};\boldsymbol{\theta}')\cdot D_{\mathrm{A}}^{2}(z_{i};\boldsymbol{\theta})}{H_{\mathrm{}}(z_{i};\boldsymbol{\theta})\cdot D_{\mathrm{A}}^{2}(z_{i};\boldsymbol{\theta}')}\right]\left[\frac{H(z_{j};\boldsymbol{\theta}')\cdot D_{\mathrm{A}}^{2}(z_{j};\boldsymbol{\theta})}{H_{\mathrm{}}(z_{j};\boldsymbol{\theta})\cdot D_{\mathrm{A}}^{2}(z_{j};\boldsymbol{\theta}')}\right]}\cdot\label{eq:obs_spectrum_cFM}\\
 &  & \cdot\left[1+\beta(z_{i};\boldsymbol{\theta}')\mu^{2}\right]\left[1+\beta(z_{j};\boldsymbol{\theta}')\mu^{2}\right]\mathscr{P}_{\mathrm{conv}}(k,z_{i},z_{j};\boldsymbol{\theta}'),\nonumber
\end{eqnarray}
 where we defined the convolved spectrum
$\mathscr{P_{\mathrm{conv}}}(k;z_{i},z_{j};\boldsymbol{\theta}')$ as
\begin{equation}
\mathscr{P_{\mathrm{conv}}}(k,z_{i},z_{j};\boldsymbol{\theta}')\equiv G(z_i;\boldsymbol{\theta}) G(z_j;\boldsymbol{\theta}) \frac{\sqrt{V_{i}V_{j}}}{(2\pi)^{3}}\int d^{3}k'P(k',z=0;\boldsymbol{\theta}')\tilde{Q}_{ij}(|\mathbf{k-k'}|)\label{eq:convolved_spectra_integral}
\end{equation}
to make the relation with Eq. (\ref{eq:obs_spectrum_lin_AP}) evident. Note that we neglected the shot noise $P_\mathrm{s}(z_i)$ because its contribution proved to be negligible with large survey bins (e.g. in the Euclid survey).

In the last two formulae, we neglected the Alcock-Paczynski (AP) effect
on the modulus $k$, the cosine $\mu$, the volumes of the shells
in $\mathscr{P_{\mathrm{conv}}}(k,z_{i},z_{j};\boldsymbol{\theta}')$ and the shell
radii $R_{i}$ entering the window functions (Eq. \ref{eq:fourier_window_anal}).
The only AP effect considered in Eq. (\ref{eq:obs_spectrum_cFM})
and (\ref{eq:convolved_spectra_integral}) is the effect on the volume
over which the power spectra are computed. In Appendix \ref{sec:The-AP-effect},
we study the complete AP effect. As explained in App. \ref{sec:The-AP-effect},
we simplified Eq. (\ref{eq:AP_final-1}) to make the computations
less time-consuming.

The convolution in Eq. (\ref{eq:corr_spectra}) is solved numerically
and the integral can also be written as a double one:
\begin{equation}
\sqrt{V_{i}V_{j}}\left\langle \delta_{k}^{(i)}\delta_{-k}^{(j)}\right\rangle =\frac{\sqrt{V_{i}V_{j}}}{(2\pi)^{3}}\int_{0}^{\infty}dk'k'^{2}P(k)\int_{-1}^{1}d\Omega\,\tilde{Q}_{ij}(\sqrt{k^{2}+k'^{2}-2kk'\Omega}).\label{eq:angular_integral}
\end{equation}
Numerical quadrature algorithms for solving integrals of this type
are time-consuming and require high precision to reach convergence.
The function $\tilde{Q}_{ij}(k)$ in the argument of the integral
is indeed a rapidly oscillating function, which is hard to integrate
accurately (Fig. \ref{fig:Q_ij plot}). A radial FFT algorithm (App.
\ref{sec:FFT-algorithm-for}) is then used for solving the radial
convolution in Eq. (\ref{eq:convolved_spectra_integral}).

As a test of the numerical accuracy, we can solve analytically the
integral in Eq. (\ref{eq:angular_integral}) by considering three
toy-model forms of power spectra. Then we compare the analytical results
with the ones obtained by using the radial FFT algorithm so that we
can calibrate the cutoff in the wavelengths and the number of samplings
(called respectively $\mathcal{R}$ and $\mathcal{N}$ in Eq. (\ref{eq:radial_FFT})
and (\ref{eq:inverse_radial_FFT}) of App. \ref{sec:FFT-algorithm-for}).

Let us first assume a constant spectrum $P(k)=\mathcal{C}$. Then the convolution
is equal to $\mathcal{C}$ for $i=j$ and is zero for $i\neq j$:
\begin{eqnarray}
\sqrt{V_{i}V_{j}}\left\langle \delta_{k}^{(i)}\delta_{-k}^{(j)}\right\rangle  & = & \frac{\mathcal{C}}{(2\pi)^{3}\sqrt{V_{i}V_{j}}}\int d^{3}k'd^{3}xd^{3}yW_{i}(|\textbf{x}|)W_{j}(|\textbf{y}|)e^{i(\mathbf{k'})\cdot(\mathbf{x+y})}e^{i(\mathbf{k})\cdot(\mathbf{x+y})}\\
 & = & \frac{\mathcal{C}}{\sqrt{V_{i}V_{j}}}\int d^{3}xd^{3}yW_{i}(|\textbf{x}|)W_{j}(|\textbf{y}|)\delta_{D}(\mathbf{x+y})e^{i(\mathbf{k})\cdot(\mathbf{x+y})}\\
 & = & \mathcal{C}\delta_{ij},\label{eq:constant_corr}
\end{eqnarray}
where we assumed that different bins do not overlap. Then we consider other two toy-model forms of spectrum:
\begin{equation}
P_{A}(k)\equiv\begin{cases}
10^{4}(\mathrm{Mpc}/h)^{3} & k\in[0.05,0.2]h/\mathrm{Mpc}\\
0 & \mathrm{otherwise}
\end{cases}\qquad P_{B}(k)\equiv\begin{cases}
10^{4}(\mathrm{Mpc}/h)^{3} & k\in[0,0.2]h/\mathrm{Mpc}\\
0 & \mathrm{otherwise}
\end{cases}\label{eq:analytical_window_spectra}
\end{equation}
In the limit $k\rightarrow0$, the double integral in Eq. (\ref{eq:angular_integral})
can be solved analytically for these two window spectra. After a proper
tuning of the parameters, we have seen that the choice of the parameters
$\mathcal{N}=7000$ and $\mathcal{R}=1.5$ $h$/Mpc leads to percentage
differences below 0.3\%, when we compare the analytical solution with
the numerical results. Thus we conclude that those values of $\mathcal{N}$
and $\mathcal{R}$ represent a good compromise between accuracy and
computational time.  Finally, Fig. \ref{fig:plot_analytical_spectrum}
shows the windowed version of the toy-model window spectrum $P_{A}(k)$
and the cross-correlation spectrum of $P_{A}(k)$ for the first and
the second Euclid bins, computed using the FFT algorithm with the
chosen parameters. We repeated the same exercise with a more realistic shape of the power spectrum $\frac{k}{1+(k/k_0)^4}$, finding similarly good results, which confirms the goodness of our choice of $\mathcal{N}$
and $\mathcal{R}$.

Now that the parameters of the radial FFT algorithm have been tuned,
they can be used to solve the convolution in Eq. (\ref{eq:convolved_spectra_integral}).
Thus we can then compute the observed cross-correlation spectra in
Eq. (\ref{eq:obs_spectrum_cFM}) and the correlated Fisher matrix
in Eq. (\ref{eq:cFM_integral}). In the next section we describe the
results for the probability distributions of the cosmological parameters.

\begin{center}
\begin{figure}
\centering
\includegraphics[scale=0.42]{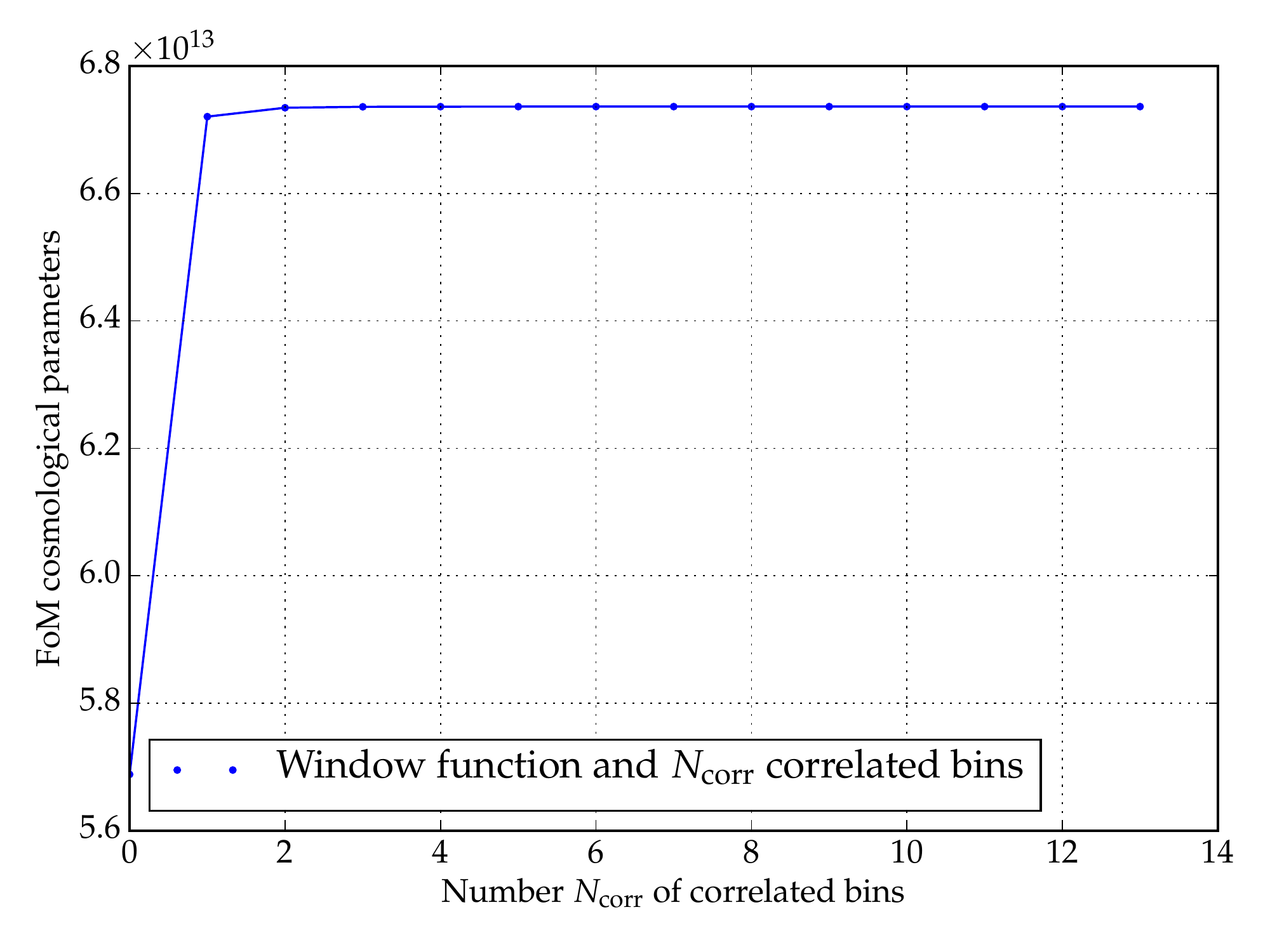}\caption{Figure of merit for all the cosmological parameters as a function
of the number $N_{\mathrm{corr}}$ of cross-correlated bins ($N_{\mathrm{corr}}=0$
means no cross-correlation, $N_{\mathrm{corr}}=1$ means that only
the cross-correlations between contiguous bins are considered, etc.).
All the bias coefficients are marginalized over. We consider the redshift
range of the Euclid mission with 14 bins (Table \ref{table:survey_specs}).
The window function effect is included. The bin redshift uncertainty
and the AP effect are not considered.}
\label{fig:FoM_correlated_bins}
\end{figure}
\end{center}

\section{Testing the assumptions: results} \label{sec:Results-and-discussion}

We will now test the assumptions regarding the effect of the window
function, the bin cross-correlations, and the bin redshift uncertainty.
In particular, we focus on the future survey of the Euclid mission.
Survey specifications are listed in Table \ref{table:survey_specs}.

Given the windowed and cross-correlation spectra computed with the
radial FFT algorithm as described in Sec. \ref{sec:computing-obs-spectr},
we analyze the probability distributions of the cosmological parameters
using the Fisher matrix approach described in Sec. \ref{sub:Fisher-information-matrix}
with the following set of cosmological parameters:
\begin{equation}
\boldsymbol{\theta}=\{h,n_{s},\Omega_{\mathrm{b}}^{(0)},\Omega_{\mathrm{cdm}}^{(0)},w_{0},\sigma_{8}\}.\label{eq:FM_parameters}
\end{equation}
The fiducial values of the cosmological parameters are taken from
the Planck results \citep{2015arXiv150201589P}. Since the bias function
is undefined, along with the parameters above, we add a bias parameter
for each bin and marginalize over it \citep{seo03}. Fiducial values
$b_{i}$ for the bias parameters are shown in Table \ref{table:survey_specs}
. The values of $k_{\mathrm{max}}$ and $k_{\mathrm{min}}$ in Eq.
(\ref{eq:uFM_integral}) and (\ref{eq:cFM_integral}) that we used
for computing the Fisher matrix are $5\cdot10^{-3}$ $h/$Mpc and
0.2 $h/$Mpc, respectively. We also experimented a value of the maximum $k$ equal to 0.15 $h/$Mpc, finding similar results to the ones we present here: the increase in the marginalized errors on the cosmological parameters averages around $6\%$, with a maximum increase of $15\%$. \textit{\emph{Our implementation of the
Fisher matrix is available publicly. The power spectra were computed
using the open source software }}\texttt{\textit{\emph{CAMB}}}\textit{\emph{
\citep{Lewis:1999bs}.}}

To analyze the probability distributions of the cosmological parameters,
we quantify: a) the fully marginalized 1-sigma uncertainties on each
of the parameters, given by $\sigma(\theta_{\alpha})=\sqrt{\left(\mathbf{F}^{-1}\right)_{\alpha\alpha}}$;
b) the marginalized confidence contour regions for pairs of parameters;
c) the figure of merit (FoM) for all the combined cosmological parameters,
defined as
\begin{equation}
\mathrm{FoM}\equiv\sqrt{\mathrm{det}\boldsymbol{F}}\label{eq:def_FOM}
\end{equation}
that is inversely proportional to the area of the marginalized confidence
contour regions.\textit{\emph{ }}

\begin{table}
\centering{}%
\begin{tabular}{c||c||>{\centering}p{19mm}|>{\centering}p{19mm}||>{\centering}p{20mm}|>{\centering}p{20mm}||>{\centering}p{21mm}|>{\centering}p{21mm}}
 &  & \multicolumn{2}{c||}{Parameters uncertainties} & \multicolumn{2}{c||}{Parameters uncertainties} & \multicolumn{2}{c}{Parameters uncertainties}\tabularnewline
 & Fiducial & \multicolumn{2}{c||}{without window function correction} & \multicolumn{2}{c||}{considering only} & \multicolumn{2}{c}{with the window function correction}\tabularnewline
 & values & \multicolumn{2}{c||}{and ignoring bin correlations} & \multicolumn{2}{c||}{the window function correction} & \multicolumn{2}{c}{and bin cross-correlations}\tabularnewline
\cline{3-8}
 &  & w/o AP & with AP & w/o AP & with AP & w/o AP & with AP\tabularnewline
\hline
\hline
$h$ & 0.67 & 0.041  & 0.026  & 0.054 (+34\%) & 0.033 (+29\%) & 0.051 (+27\%)  & 0.032 (+23\%) \tabularnewline
$n_{s}$ & 0.965 & 0.035  & 0.020 & 0.045 (+29\%) & 0.024 (+22\%) & 0.043 (+23\%)  & 0.023 (+17\%) \tabularnewline
$\Omega_{\mathrm{b}}^{(0)}$ & 0.049 & 0.0030  & 0.0020 & 0.0042 (+38\%) & 0.0027 (+37\%) & 0.0039 (+30\%)  & 0.0026 (+29\%) \tabularnewline
$\Omega_{\mathrm{cdm}}^{(0)}$ & 0.2673 & 0.0026  & 0.0040 & 0.0034 (+31\%)  & 0.0046 (+15\%) & 0.0033 (+26\%)  & 0.0045 (+12\%) \tabularnewline
$w_{0}$ & -0.98 & 0.043  & 0.022 & 0.044 (+1.0\%)  & 0.023 (+6.5\%)  & 0.044 (+1\%)  & 0.0233 (+5.7\%) \tabularnewline
$\sigma_{8}$ & 0.83 & 0.0092 & 0.0021 & 0.0103 (+11\%) & 0.0021 (+0\%) & 0.0101 (+9\%) & 0.0021 (+0\%)\tabularnewline
\end{tabular}\caption{For each parameter, the table contains the fiducial values of the
cosmological parameters (column one), the values of the marginalized
1-sigma uncertainties computed without neither window function effect
nor bin correlations (columns two and three), the values of the uncertainties
computed with only the window function (columns four and five), and
the values of the uncertainties computed with the window function
and bin correlations between contiguous bins (columns six and seven).
In columns four to seven we also provide the percentages of the differences
of the corresponding values to the values in columns two and three.
All computations were done with and without considering the AP effect
(Sec. \ref{sec:computing-obs-spectr} and App. \ref{sec:The-AP-effect})
and with respect to the Euclid specifications (Table \ref{table:survey_specs}). }
\label{table:percentage_difference}
\end{table}
\begin{table}
\centering{}%
\begin{tabular}{c||c||>{\centering}p{20mm}|>{\centering}p{20mm}||>{\centering}p{20mm}|>{\centering}p{20mm}||>{\centering}p{20mm}|>{\centering}p{20mm}}
 &  & \multicolumn{2}{c||}{Parameters uncertainties} & \multicolumn{2}{c||}{Parameters uncertainties} & \multicolumn{2}{c}{Parameters uncertainties}\tabularnewline
 & Fiducial & \multicolumn{2}{c||}{without window function correction} & \multicolumn{2}{c||}{considering only} & \multicolumn{2}{c}{with the window function correction}\tabularnewline
 & values & \multicolumn{2}{c||}{and ignoring bin correlations} & \multicolumn{2}{c||}{the window function correction} & \multicolumn{2}{c}{and bin cross-correlations}\tabularnewline
\cline{3-8}
 &  & w/o redshift parameters & with redshift parameters & w/o redshift parameters & with redshift parameters & w/o redshift parameters & with redshift parameters\tabularnewline
\hline
\hline
$h$ & 0.67 & 0.041  & 0.041 (-\%)  & 0.054 & 0.054 (-\%)  & 0.051 & 0.051 (-\%) \tabularnewline
$n_{s}$ & 0.965 & 0.035  & 0.035 (-\%)  & 0.045 & 0.046 (-\%)  & 0.043 & 0.043 (-\%) \tabularnewline
$\Omega_{\mathrm{b}}^{(0)}$ & 0.049 & 0.0030  & 0.0030 (-\%)  & 0.0042 & 0.0042 (-\%)  & 0.0039 & 0.0039 (-\%) \tabularnewline
$\Omega_{\mathrm{cdm}}^{(0)}$ & 0.2673 & 0.0026  & 0.0026 (-\%)  & 0.0034 & 0.0034 (-\%)  & 0.0033 & 0.0033 (-\%) \tabularnewline
$w_{0}$ & -0.98 & 0.043  & 0.051 (+18\%)  & 0.044  & 0.051 (+17\%)  & 0.044 & 0.051 (+17\%) \tabularnewline
$\sigma_{8}$ & 0.83 & 0.0092 & 0.011 (+14\%) & 0.0103  & 0.012 (+12\%) & 0.0101 & 0.011 (+12\%)\tabularnewline
\end{tabular}\caption{For each parameter, the table contains the fiducial values of the
cosmological parameters (column one), the values of the marginalized
1-sigma uncertainties computed without neither window function effect
nor bin correlations (columns two and three), the values of the uncertainties
computed with only the window function (columns four and five), and
the values of the uncertainties computed with the window function
and bin correlations between contiguous bins (columns six and seven).
In columns three, five and seven we also provide the percentages of
the differences between the values with and without bin redshift uncertainty.
All computations were done without considering the AP effect (Sec.
\ref{sec:computing-obs-spectr} and App. \ref{sec:The-AP-effect})
and with respect to the Euclid specifications (Table \ref{table:survey_specs}). }
\label{table:percentage_difference_redshift}
\end{table}

\subsection{Euclid survey} \label{sub:Testing-the-assumptions}

Since computing bin correlations for each pair of bins is computationally
expensive, particularly with the complete AP effect (App. \ref{sec:The-AP-effect}),
it is desirable to reduce the number of considered pairs. We tested
how good is the approximation to consider only the correlations between
contiguous bins. In Fig. \ref{fig:FoM_correlated_bins}, we plot the
FoM values computed with bin correlations for neighboring pairs of
bins, so that the number of neighbors $N_{\mathrm{corr}}$ is given
by the horizontal axis, i.e. if the number of correlated bins is equal
to one then bin correlations are computed for contiguous bins only;
if the number of correlated bins is equal to 13 then bin correlations
are computed for all pairs of bins. The figure shows that the FoM
values does not increase with an increasing number of bins beyond
the nearest neighbor when the bin size is $\Delta z=0.1$. Thus, we
can conclude that it is sufficient to focus on nearby bins and in
the following we consider correlations only among bins closer than
$\Delta z=0.1$. Note that in our analysis we neglected the lensing effect. By adding it, we expect the correlation terms to become more important and we are currently working on it. As shown in \cite{montanari2015measuring} (Fig. 2), the lensing effect modifies mainly the correlations between distant bins, which in our analysis are neglected because of their small contribution to the final constraints.

For each parameter, Table \ref{table:percentage_difference_redshift}
contains the fiducial values of the cosmological parameters (column
one), the values of the marginalized 1-sigma uncertainties computed
without neither window function effect nor bin correlations (columns
two and three), the values of the uncertainties computed with only
the window function (columns four and five), and the values of the
uncertainties computed with the window function and bin correlations
(columns six and seven). In columns four to seven we also provide
the percentages of the differences of the corresponding values to
the values in columns two and three. All computations were done with
and without considering the AP effect (Sec. \ref{sec:computing-obs-spectr}
and App. \ref{sec:The-AP-effect}) and with respect to the Euclid
specifications (Table \ref{table:survey_specs}).

As shown in Table \ref{table:percentage_difference_redshift}, the
window function has a significant effect (15-38\% difference from
the case without the window function) on parameters $h,n_{s},\Omega_{\mathrm{b}}^{(0)},\Omega_{\mathrm{cdm}}^{(0)}$.
Table \ref{table:percentage_difference_redshift} also shows that
adding bin correlations reduces the uncertainties for all considered
parameters as compared to the case with only the window function.
There is still a significant difference on parameters
$h,n_{s},\Omega_{\mathrm{b}}^{(0)},\Omega_{\mathrm{cdm}}^{(0)}$ (12-30\%)
as compared to the case without neither the window function nor bin
correlations.

\begin{figure}
\centering
\includegraphics[scale=0.52]{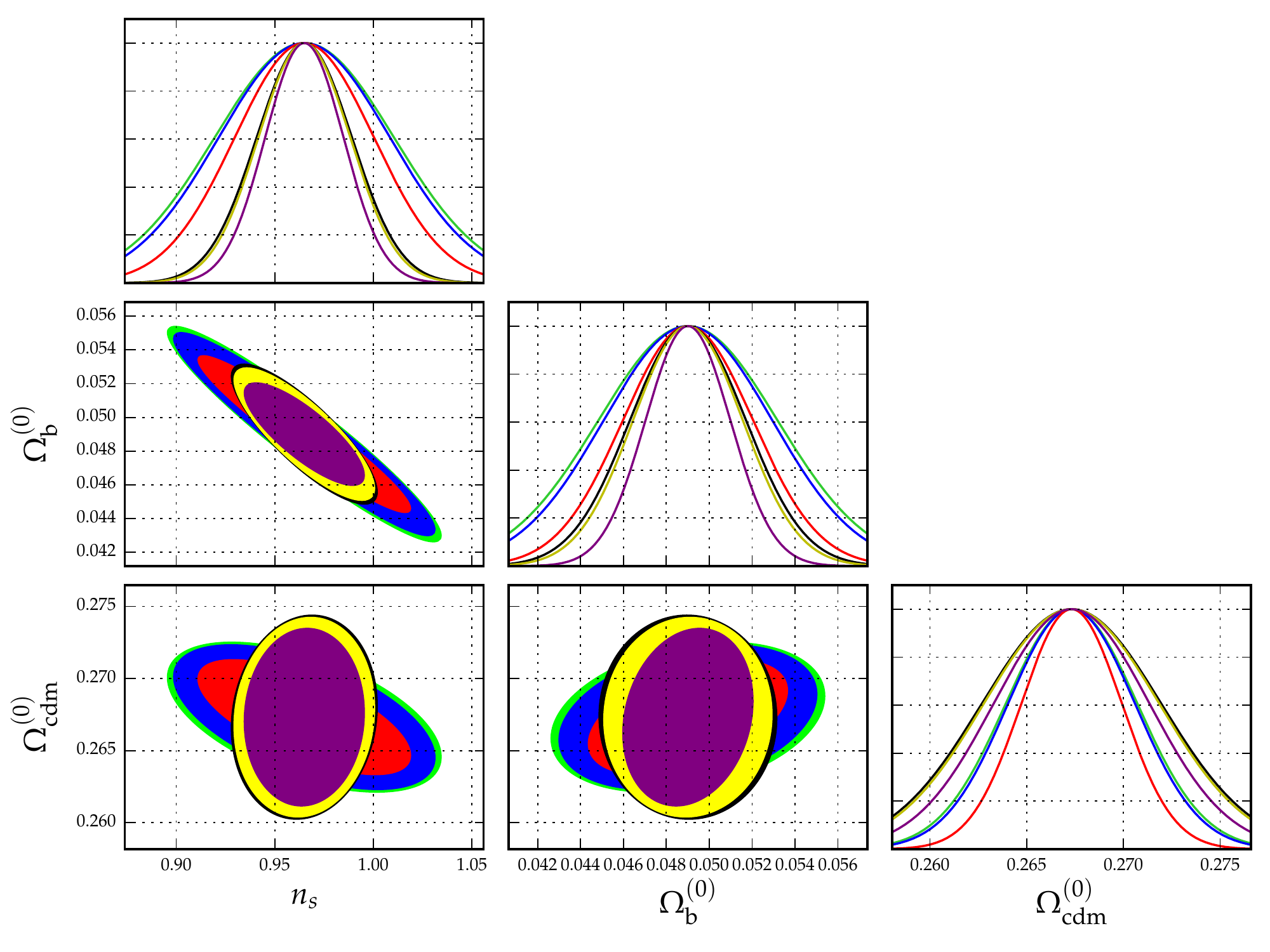}\caption{Marginalized confidence contour regions (1-sigma contours) for pairs
of cosmological parameters $n_{s}$, $\Omega_{\mathrm{b}}^{(0)}$
and $\Omega_{\mathrm{cdm}}^{(0)}$ with and without the Alcock-Paczynski
effect computed towards the Euclid specifications (Table \ref{table:survey_specs}).
The red, green and blue ellipses are computed without AP effect (Sec.
\ref{sec:computing-obs-spectr} and App. \ref{sec:The-AP-effect}).
The purple, black and yellow ellipses are computed with the AP effect.
The red and purple ellipses are computed without neither window function
nor bin correlations, the green and black ones include the effect
of the window function, the blue and yellow ones include both the
correlations and the window function. We consider only correlations
between contiguous bins. We also show the one-dimensional probability
distributions for each considered parameter.}
\label{fig:ellipses_AP_effect}
\end{figure}
\begin{figure}
\centering
\includegraphics[scale=0.73]{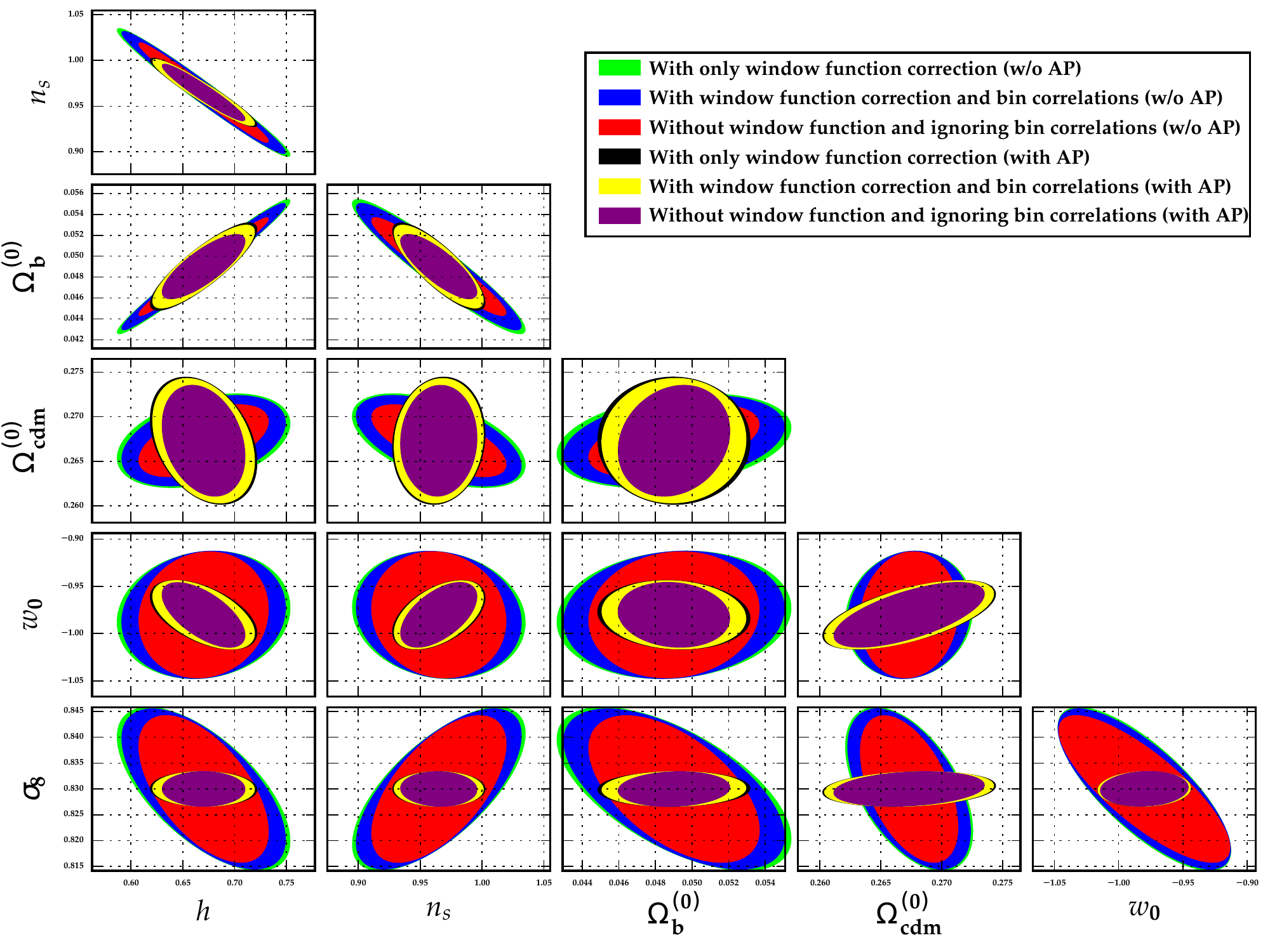}\caption{As previous Fig. \ref{fig:ellipses_AP_effect}, but for pairs of all
the considered cosmological parameters. The one-dimensional probability
distributions are omitted.}
\label{fig:ellipses_AP_all}
\end{figure}
\begin{figure}
\centering
\includegraphics[scale=0.65]{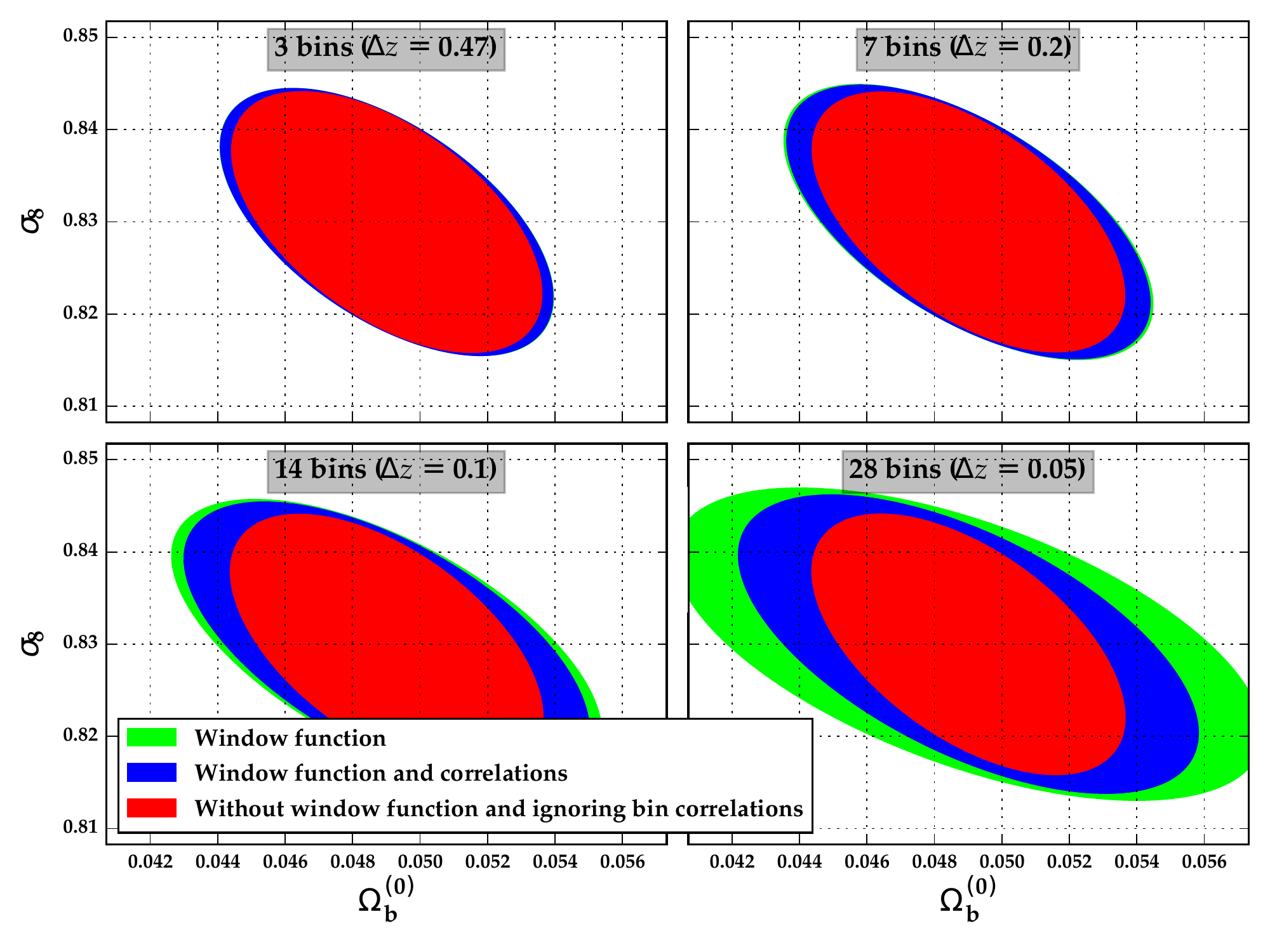}\caption{Marginalized confidence contour regions for $\sigma_{8}$ and $\Omega_{\mathrm{b}}^{(0)}$
(1-sigma contours) for different numbers of bins. The red ellipses
are computed using a Fisher matrix approach without considering the
window function effect and cross-correlations between bins. The green
ellipses are computed with the window function. The blue ellipses
are computed with the window function and correlations between bins.
The AP effect is neglected.}
\label{fig:ellipses_widthbins}
\end{figure}

Table \ref{table:percentage_difference_redshift} shows that the complete
AP effect has a considerable influence on the uncertainties of parameters
$\Omega_{\mathrm{cdm}}^{(0)},w_{0},\sigma_{8}$. The results are also
shown in Fig. \ref{fig:ellipses_AP_effect} and \ref{fig:ellipses_AP_all}.
As explained in App. \ref{sec:The-AP-effect}, we simplified Eq. (\ref{eq:AP_final-1})
to make the computations less time-consuming. Even with the simplified
Eq. (\ref{eq:AP_final-1}), the processing runtime is orders of magnitude
larger as compared to the runtime without the AP effect and by using
FFT algorithms. Thus, ignoring the AP effect significantly reduces
the runtime when the window function or bin correlations are considered.

In Table \ref{table:percentage_difference_redshift} we consider the
redshift of the bins as an additional parameter. The differences with
the values computed without considering the redshift as an additional
parameter are significant only for the parameters $\sigma_{8}$ and
$w_{0}$ (12-18\% difference).

\begin{figure}
\begin{minipage}[t]{0.48\columnwidth}%
\begin{center}
\includegraphics[scale=0.42]{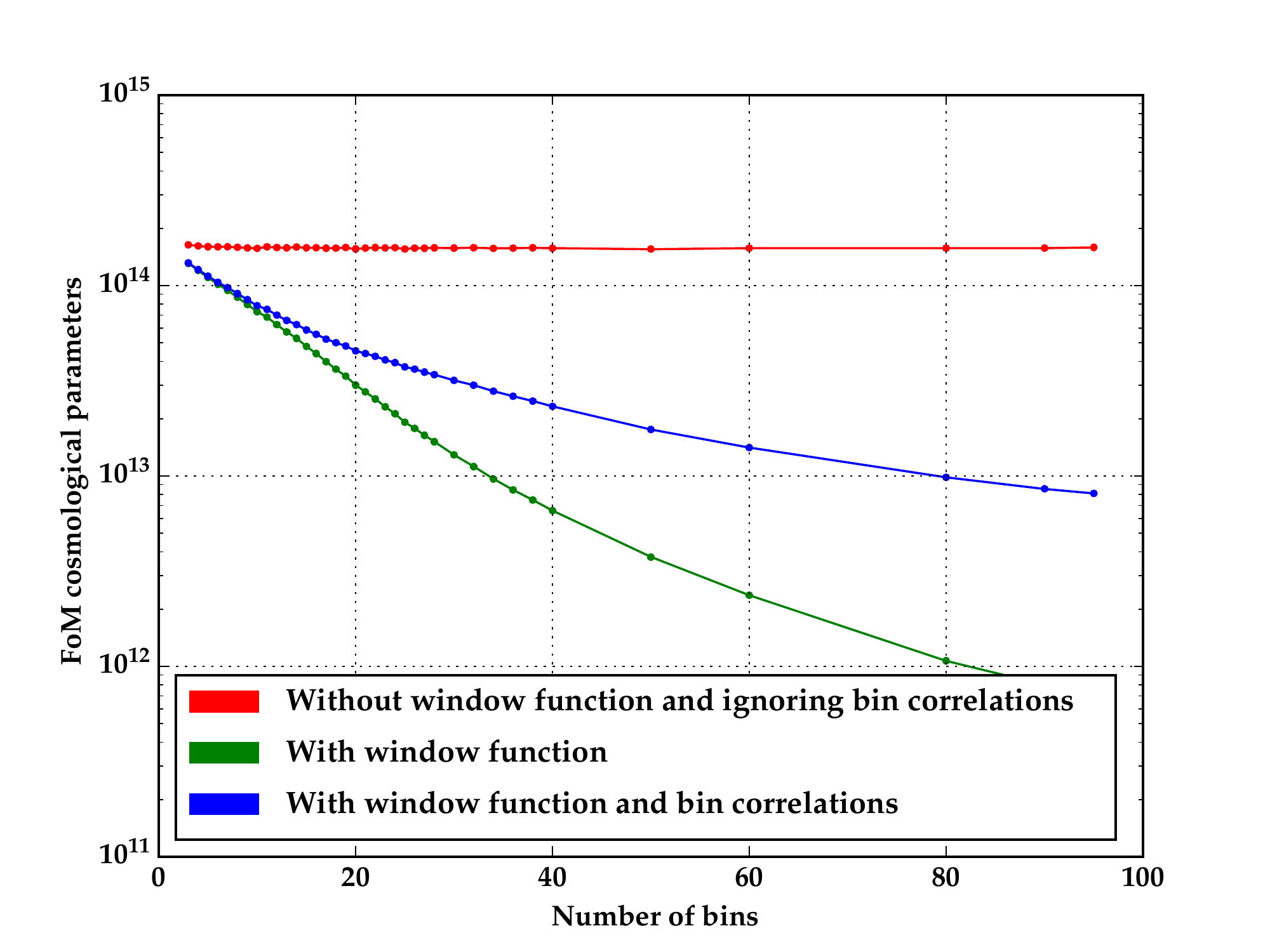}\caption{Figure of merit for all the cosmological parameters as a function
of the number of bins. All the bias coefficients are marginalized
over. We consider the redshift range of the Euclid mission. The red
plot is computed without considering neither the window function effect
nor cross-correlations between bins. The green plot is computed with
just the window function effect and the blue plot is with the window
function effect and bin correlations. For the case with correlations
(blue plot) we compute bin correlations only for the bin pairs such
that the distance between the bins is less than $\Delta z=$0.1}
\label{fig:FOM_plot}
\par\end{center}%
\end{minipage}\hfill{}%
\begin{minipage}[t]{0.48\columnwidth}%
\begin{center}
\includegraphics[scale=0.42]{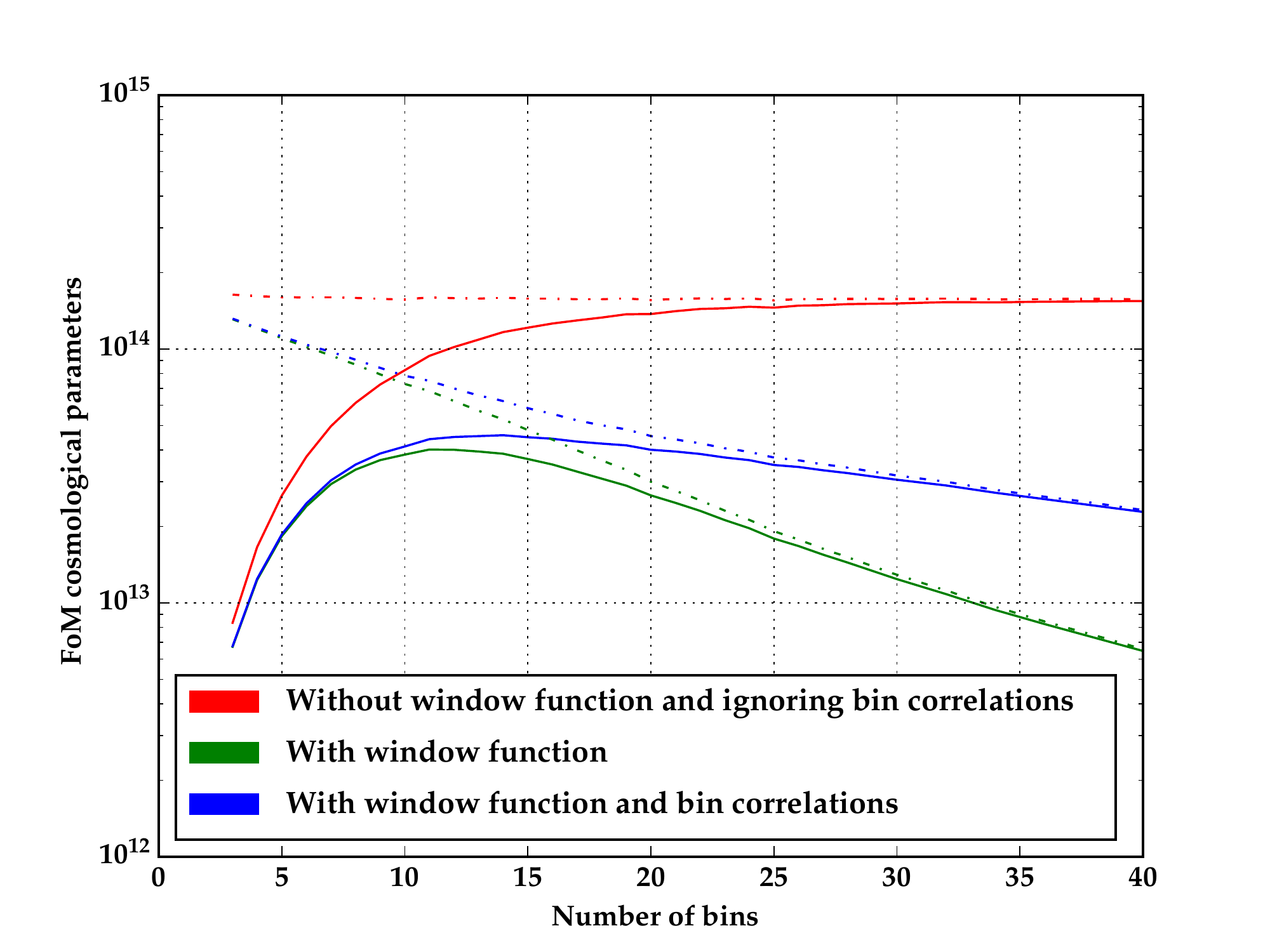}\caption{As previous Fig. \ref{fig:FOM_plot}, now including the bin redshift
uncertainty (solid lines), compared with the case without (dashed
lines). }
\label{fig:FOM_redshift_parameter}
\par\end{center}%
\end{minipage}
\end{figure}
\begin{figure}
\centering
\includegraphics[scale=0.42]{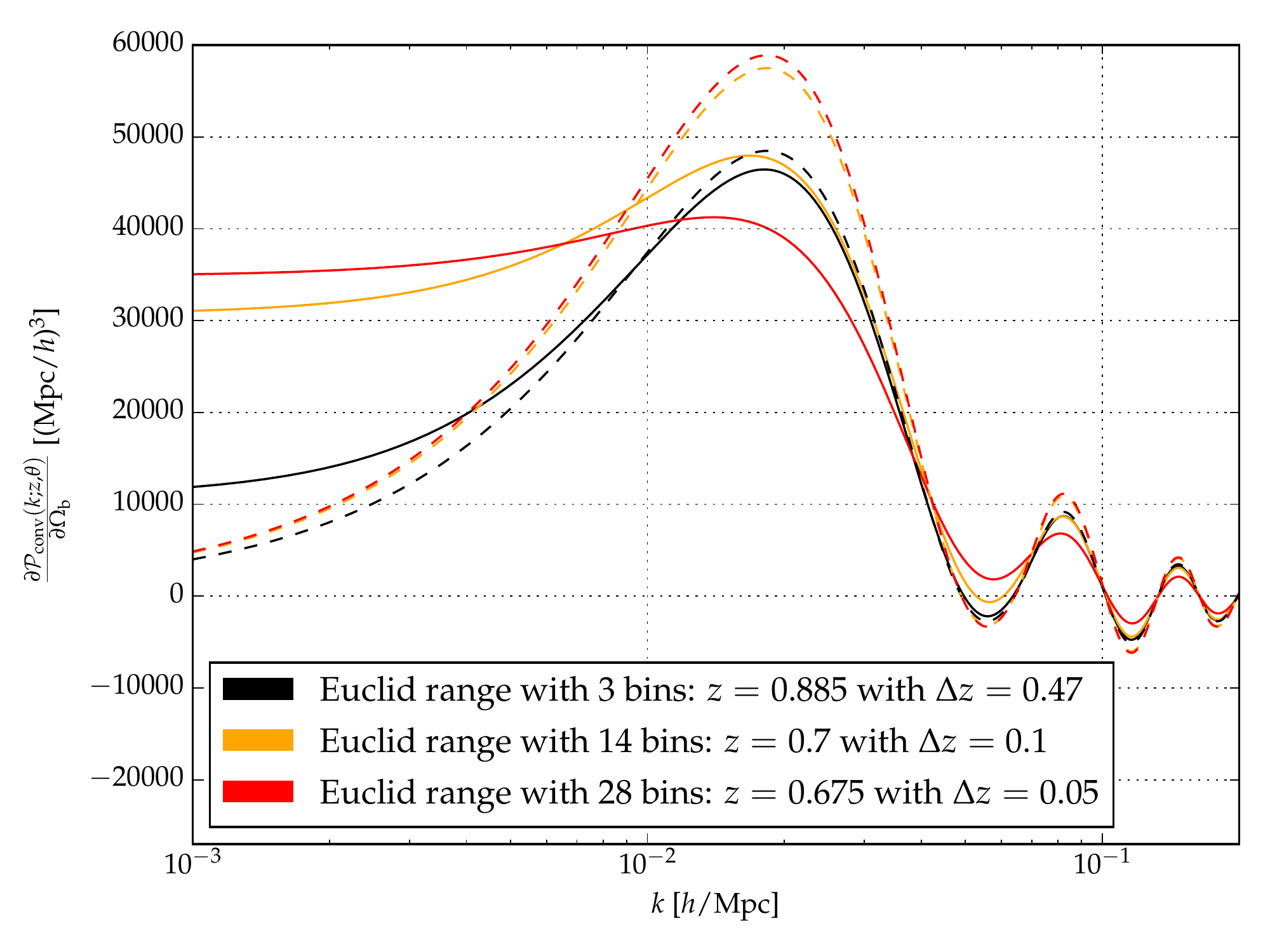}\caption{Derivatives with respect to parameter $\Omega_{\mathrm{b}}^{(0)}$
of the convolved spectrum in Eq. (\ref{eq:convolved_spectra_integral})
(solid lines) and the standard linear spectrum in Eq. (\ref{eq:lin_spectrum})
(dashed lines) for different numbers of bins. We consider the redshift
range of the Euclid mission. We subdivide this range into 3, 14 and
28 bins (black, orange, red plots, respectively). Note that for increasing
number of bins, in the region of $k\in[10^{-2},10^{-1}]$ $h$/Mpc,
the values of the convolved derivatives are smaller with respect to
the derivatives of the standard linear spectra.}
\label{fig:conv_1}
\end{figure}

\subsection{Varying number of bins} \label{sub:Assumption-effects-with}

We test the effects of the assumptions with respect to the number
of bins. We consider the redshift range of the Euclid mission, $z\in[0.65,2.05]$,
see Table \ref{table:survey_specs}. We subdivide this range into
$N$ bins and vary $N$ in range from 2 to 95. The galaxy density $n(z_i)$ at a redshift $z_i$ is still assumed to be constant inside the bin. For a number of bins different from the standard 14 bins considered for the Euclid mission, the values $n(z_i)$ and $b(z_i)$ are found by performing a fit of the values shown in Table \ref{table:survey_specs}. As explained in Sec.
\ref{sub:Testing-the-assumptions}, since computing bin correlations
for each pair of bins (here max. 4465 pairs) is computationally expensive,
it is desirable to reduce the number of considered pairs. For Euclid
with 14 bins, we concluded that it is sufficient to focus on the contiguous
bins. Since the width of a Euclid bin is equal to $\Delta z=0.1$,
in the following we compute always correlations for contiguous bins
if the distance between bin boundaries is $\Delta z\ge0.1$ and among
all the bins closer than $\Delta z=0.1$ if the bins are smaller than
$0.1$. Then for instance with $N=20$, the distance between two contiguous
bins is $\Delta z_{\mathrm{bins}}=0.07$, thus we consider correlations
between bins $i,i\pm1,i\pm2$.

Figure \ref{fig:FOM_plot} shows the dependence of the FoM (Eq. \ref{eq:def_FOM})
on the number of bins for all combined cosmological parameters in
Eq. (\ref{eq:FM_parameters}). The red plot is computed using the
Fisher matrix without considering neither the window function effect
nor cross-correlations between bins. The green plot is computed with
just the window function effect and the blue plot is computed with
the window function effect and bin correlations. Figure \ref{fig:ellipses_widthbins}
shows the marginalized confidence contour regions for $\sigma_{8}$
and $\Omega_{\mathrm{b}}^{(0)}$ for different numbers of bins. The
other parameters are marginalized over. The red ellipses are computed
using the Fisher matrix without considering the window function effect
and cross-correlations between bins. The green ellipses are computed
with only the window function. The blue ellipses are computed with
the window function and correlations between bins. Based on these
figures, we analyze the effects of the window function and the bin
correlations. In both Fig. \ref{fig:ellipses_widthbins} and \ref{fig:FOM_plot},
redshift is not considered as additional parameter, thus there is
no uncertainty on the value of the redshift bins. \\

\textsc{No bin cross-correlations, no window function effect}. As
mentioned in Sec. \ref{sub:Fisher-information-matrix}, without bin
cross-correlations, the random variables assigned to Fourier coefficients
$\{\delta_{\mathbf{k}}^{(1)},...\,,\delta_{\mathbf{k}}^{(N)}\}$ are
statistically independent. If we increase $N$, one would expect smaller
uncertainties on the parameters $\boldsymbol{\theta}$ (Eq. \ref{eq:FM_parameters}).
However, Fig. \ref{fig:FOM_plot} (red curve) shows that the value
of the FoM (Eq. \ref{eq:def_FOM}) does not significantly vary with
increasing number of bins, because the volume factor $V_{\mathrm{i}}$
(Eq. \ref{eq:eff_volume}) compensates the increased value of $N$.\footnote{Note that the Fisher matrix element (Eq. \ref{eq:uFM_integral}) does
not significantly vary, but it is not constant. The reason is that
all the spectra are computed at the median bin redshifts and these
values vary for different numbers of bins.} In other words, we consider more bins, but the amount of information
we obtain from each of them is less. Similarly in Fig. \ref{fig:ellipses_widthbins}
red ellipses have a similar shape for different number of bins. \\

\textsc{Window function effect without bin cross-correlations.} We
now test the effects of the window function and consider the observed
convolved power spectrum in Eq. (\ref{eq:obs_spectrum_cFM}) for $i=j$.
Figure \ref{fig:ellipses_widthbins} shows that green ellipses are
larger for larger numbers of bins. Similarly, the FoM in Fig. \ref{fig:FOM_plot}
(green plot) decreases with an increasing number of bins. Thus, the
window function results in larger uncertainties for larger numbers
of bins.

This effect can be also seen in Fig. \ref{fig:conv_1}, which shows
the derivatives with respect to parameter $\Omega_{\mathrm{m}}^{(0)}$
of the convolved spectrum (Eq. \ref{eq:convolved_spectra_integral})
and the standard linear spectrum (Eq. \ref{eq:lin_spectrum}) for
different numbers of bins. Note that for an increasing number of bins,
in the region of $k\in[10^{-2},10^{-1}]$ $h$/Mpc, the values of
the convolved derivatives are smaller with respect to the derivatives
of the standard linear spectra. Thus, with the window function effect
the spectrum derivatives in Eq. (\ref{eq:uFM_integral}) have smaller
values and a Fisher matrix with smaller values corresponds to larger
uncertainties.\\

\textsc{Bins cross-correlations and window function effect.} We now
test the second assumption and consider correlations between different
bins. Figure \ref{fig:ellipses_widthbins} shows that uncertainties
with only the window function (green ellipses) are larger as compared
to the ones with bin correlations and window function (blue ellipses).
The figure also shows that adding bin correlations decreases the uncertainties
for a larger number of bins as compared to the case with only the
window function. The same effect can be seen in Fig. \ref{fig:FOM_plot},
where the FoM values with only window function (green plot) decrease
more as compared to the values with both the window function and bin
correlations (blue plot). In the case with both the window function
and bin correlations, note that the values of the FoM (blue plot)
decrease slowly for larger numbers of bins, but the FoM does not reach
a constant value as one would expect. We conjecture that, for a large
number of bins, neglecting the mode-mode correlations for $\boldsymbol{k\neq k'}$
decreases the values of the FoM. \\

\textsc{Redshift as additional parameter}. In Subsection \ref{sub:Redshift-as-an},
we considered the redshift of the bins as an additional parameter.
Figure \ref{fig:FOM_redshift_parameter} shows the FoM computed by
using Fisher matrix $\hat{F}_{AB}$ (Sec. \ref{sub:Redshift-as-an})
marginalized over redshift parameters $z_{i}$ and bias parameters
$b_{i}$. Note that for a small number of bins the values of the FoM
are smaller as compared to the values of the FoM shown in Fig. \ref{fig:FOM_plot}.
Thus, if we consider the redshifts of the bins as parameters, this
results in larger uncertainties on the cosmological parameters for
smaller numbers of bins. In the case with the window function and
bin correlations the values of the uncertainties present a maximum
for 14 bins, that is usually the number considered for the Euclid
survey (Table \ref{table:survey_specs}).\textbf{ }For $N\apprge25$,
the values of the FoM computed with and without redshift as additional
parameter are close to each other, i.e. the addition of the bin-redshift
uncertainty do not modify the results. We also experimented with another
value of the standard deviation $\sigma_{i}$ (Eq. \ref{eq:uncorr_likelihood_redshift})
equal to half of the bin width. With this increased bin redshift uncertainty,
the values of the FoM are smaller than compared to the one shown in
Fig. \ref{fig:FOM_redshift_parameter} and in the case with the window
function and bin correlations (blue plot) the maximal value of the
FoM is reached around \textasciitilde{}25 bins.

\section{Conclusions} \label{sec:Conclusions}

Forecast studies for galaxy clustering mostly rely on the Fisher matrix
method and typically imply several assumptions, including neglecting
the window function effect and the bin cross-correlation spectra and
fixing the bin redshift range to the median value. In this article
we proposed an approach for testing if these assumptions hold for
realistic surveys, in particular the Euclid survey and we estimated
the change on the parameter forecast. We also investigated the dependency
of the effects on the number of redshift bins and included the Alcock-Paczynski
effect in a simplified form. For computations we used an FFT algorithm
and implemented an optimized Fisher matrix. The main results of the
article are summarized in Table \ref{table:percentage_difference}
and \ref{table:percentage_difference_redshift}. The results suggest
that the window function and the bin cross-correlations, although
acting in opposite sense, have a considerable combined influence on
the forecasted errors for a Euclid-like survey, amounting to 10-30\%
for several of the cosmological parameters.

Possible improvements of our approach that we will pursue in future
work include the following. First, the dependence of the survey volume
and $k_{\mathrm{max}}$ and $k_{\mathrm{min}}$ in Eq. (\ref{eq:uFM_integral}) on the
cosmological parameters can be considered. A second possible improvement
concerns a further optimization of the algorithm for computing the
derivatives of the convolved spectra with the Alcock-Paczynski effect,
which proved to be computationally expensive in our study. Third,
further models besides $\Lambda$CDM should be considered. Fourth,
whereas we considered only the linear power spectrum, non-linearities
could also taken into consideration. Fifth, we plan to assess the
impact of the mode-mode correlation, so far neglected in most analyses.
We expect it to be particularly important for small bins. Finally,
other approximations that need to be tested and to be improved upon
are the flat-sky redshift distortion and the neglect of large-scale
relativistic effects like lensing.

\section*{ACKNOWLEDGMENTS }

We would like to thank Katya Ovchinnikova and Santiago Casas for their
various contributions that made this article possible and Ruth Durrer, Daniele
Bertacca for useful discussions. This research has been supported
by DFG through the grant TRR33 ``The Dark Universe". ASM acknowledges
financial support from the graduate college \emph{Astrophysics of
cosmological probes of gravity} by Landesgraduiertenakademie Baden-W{\"u}rttemberg.




\bibliographystyle{mnras}
\bibliography{Improving_galaxy_clustering} 

\begin{thebibliography}{}
\makeatletter
\relax
\def\mn@urlcharsother{\let\do\@makeother \do\$\do\&\do\#\do\^\do\_\do\%\do\~}
\def\mn@doi{\begingroup\mn@urlcharsother \@ifnextchar [ {\mn@doi@}
  {\mn@doi@[]}}
\def\mn@doi@[#1]#2{\def\@tempa{#1}\ifx\@tempa\@empty \href
  {http://dx.doi.org/#2} {doi:#2}\else \href {http://dx.doi.org/#2} {#1}\fi
  \endgroup}
\def\mn@eprint#1#2{\mn@eprint@#1:#2::\@nil}
\def\mn@eprint@arXiv#1{\href {http://arxiv.org/abs/#1} {{\tt arXiv:#1}}}
\def\mn@eprint@dblp#1{\href {http://dblp.uni-trier.de/rec/bibtex/#1.xml}
  {dblp:#1}}
\def\mn@eprint@#1:#2:#3:#4\@nil{\def\@tempa {#1}\def\@tempb {#2}\def\@tempc
  {#3}\ifx \@tempc \@empty \let \@tempc \@tempb \let \@tempb \@tempa \fi \ifx
  \@tempb \@empty \def\@tempb {arXiv}\fi \@ifundefined
  {mn@eprint@\@tempb}{\@tempb:\@tempc}{\expandafter \expandafter \csname
  mn@eprint@\@tempb\endcsname \expandafter{\@tempc}}}

\bibitem[\protect\citeauthoryear{Abell et~al.,}{Abell
  et~al.}{2009}]{abell2009lsst}
Abell P.~A.,  et~al., 2009, preprint (\mn@eprint {arXiv} {0912.0201})

\bibitem[\protect\citeauthoryear{Albrecht et~al.,}{Albrecht
  et~al.}{2009}]{albrecht2009findings}
Albrecht A.,  et~al., 2009, preprint (\mn@eprint {arXiv} {0901.0721})

\bibitem[\protect\citeauthoryear{Alcock \& Paczy\'nski}{Alcock \&
  Paczy\'nski}{1979}]{Alcock:1979mp}
Alcock C.,  Paczy\'nski B.,  1979, Nature, 281, 358

\bibitem[\protect\citeauthoryear{Amendola \& Tsujikawa}{Amendola \&
  Tsujikawa}{2010}]{Amendola2010}
Amendola L.,  Tsujikawa S.,  2010, Dark Energy: Theory and Observations.
Cambridge University Press, Cambridge; New York

\bibitem[\protect\citeauthoryear{Amendola, Quercellini  \& Giallongo}{Amendola
  et~al.}{2005}]{2005MNRAS.357..429A}
Amendola L.,  Quercellini C.,   Giallongo E.,  2005, \mn@doi [Mon. Not. R.
  Astron. Soc.] {10.1111/j.1365-2966.2004.08558.x}, \href
  {http://adsabs.harvard.edu/abs/2005MNRAS.357..429A} {357, 429}

\bibitem[\protect\citeauthoryear{Amendola et~al.,}{Amendola
  et~al.}{2013}]{1206.1225}
Amendola L.,  et~al., 2013, \mn@doi [Living Rev. Relativity 16, (2013), 6]
  {10.12942/lrr-2013-6}

\bibitem[\protect\citeauthoryear{Baumgart \& Fry}{Baumgart \&
  Fry}{1991}]{baumgart1991fourier}
Baumgart D.~J.,  Fry J.,  1991, The Astrophysical Journal, 375, 25

\bibitem[\protect\citeauthoryear{Bonaldi, Harrison, Camera  \& Brown}{Bonaldi
  et~al.}{2016}]{bonaldi2016ska}
Bonaldi A.,  Harrison I.,  Camera S.,   Brown M.~L.,  2016, preprint
  (\mn@eprint {arXiv} {1601.03948})

\bibitem[\protect\citeauthoryear{{DESI Collaboration}}{{DESI
  Collaboration}}{2016}]{DESI}
{DESI Collaboration} 2016, The {Dark Energy Spectroscopic Instrument (DESI)},
  \url{http://desi.lbl.gov/}

\bibitem[\protect\citeauthoryear{Di~Porto, Amendola  \& Branchini}{Di~Porto
  et~al.}{2012}]{di2012simultaneous}
Di~Porto C.,  Amendola L.,   Branchini E.,  2012, Monthly Notices of the Royal
  Astronomical Society: Letters, 423, L97

\bibitem[\protect\citeauthoryear{Dodelson}{Dodelson}{2003}]{dodelson2003modern}
Dodelson S.,  2003, Modern cosmology.
Academic press

\bibitem[\protect\citeauthoryear{{Feldman}, {Kaiser}  \& {Peacock}}{{Feldman}
  et~al.}{1994}]{FKP}
{Feldman} H.~A.,  {Kaiser} N.,   {Peacock} J.~A.,  1994, \mn@doi [\apj]
  {10.1086/174036}, \href {http://adsabs.harvard.edu/abs/1994ApJ...426...23F}
  {426, 23}

\bibitem[\protect\citeauthoryear{Fisher}{Fisher}{1935}]{fisher1935logic}
Fisher R.~A.,  1935, Journal of the royal statistical society, 98, 39

\bibitem[\protect\citeauthoryear{Fisher, Scharf  \& Lahav}{Fisher
  et~al.}{1994}]{fisher1994spherical}
Fisher K.~B.,  Scharf C.~A.,   Lahav O.,  1994, Monthly Notices of the Royal
  Astronomical Society, 266, 219

\bibitem[\protect\citeauthoryear{Flaugher}{Flaugher}{2005}]{collaboration2005dark}
Flaugher B.,  2005, International Journal of Modern Physics A, 20, 3121

\bibitem[\protect\citeauthoryear{Groth \& Peebles}{Groth \&
  Peebles}{1977}]{groth1977statistical}
Groth E.~J.,  Peebles P.,  1977, The Astrophysical Journal, 217, 385

\bibitem[\protect\citeauthoryear{Heavens}{Heavens}{2009}]{heavens2009statistical}
Heavens A.,  2009, preprint (\mn@eprint {arXiv} {0906.0664})

\bibitem[\protect\citeauthoryear{Heavens \& Taylor}{Heavens \&
  Taylor}{1995}]{heavens1995spherical}
Heavens A.,  Taylor A.,  1995, Monthly Notices of the Royal Astronomical
  Society, 275, 483

\bibitem[\protect\citeauthoryear{Hill et~al.,}{Hill et~al.}{2008}]{0806.0183}
Hill G.~J.,  et~al., 2008, ASP Conf.Ser.399:115-118,2008

\bibitem[\protect\citeauthoryear{Kaiser et~al.,}{Kaiser
  et~al.}{2002}]{kaiser2002pan}
Kaiser N.,  et~al., 2002, in Astronomical Telescopes and Instrumentation. pp
  154--164

\bibitem[\protect\citeauthoryear{Lahav, Lilje, Primack  \& Rees}{Lahav
  et~al.}{1991}]{Lahav-etal:1991}
Lahav O.,  Lilje P.,  Primack J.,   Rees M.,  1991, Mon. Not. R. Astron. Soc.,
  \href {http://adsabs.harvard.edu/abs/1991MNRAS.251..128L} {251, 128}

\bibitem[\protect\citeauthoryear{Laureijs et~al.}{Laureijs
  et~al.}{2011}]{euclidredbook}
Laureijs R.,  et~al., 2011, Technical Report ESA/SRE(2011)12, Euclid: Mapping
  the geometry of the dark Universe. {Definition} Study Report.
Paris (\mn@eprint {arXiv} {1110.3193})

\bibitem[\protect\citeauthoryear{Lewis, Challinor  \& Lasenby}{Lewis
  et~al.}{2000}]{Lewis:1999bs}
Lewis A.,  Challinor A.,   Lasenby A.,  2000, \mn@doi [Astrophys. J.]
  {10.1086/309179}, 538, 473

\bibitem[\protect\citeauthoryear{Linder}{Linder}{2005}]{Linder:2005in}
Linder E.,  2005, \mn@doi [Phys. Rev. D] {10.1103/PhysRevD.72.043529}, 72

\bibitem[\protect\citeauthoryear{Lyth \& Liddle}{Lyth \&
  Liddle}{2009}]{lyth2009primordial}
Lyth D.~H.,  Liddle A.~R.,  2009, The primordial density perturbation:
  Cosmology, inflation and the origin of structure.
Cambridge University Press

\bibitem[\protect\citeauthoryear{Montanari \& Durrer}{Montanari \&
  Durrer}{2012}]{montanari2012new}
Montanari F.,  Durrer R.,  2012, Physical Review D, 86, 063503

\bibitem[\protect\citeauthoryear{Montanari \& Durrer}{Montanari \&
  Durrer}{2015}]{montanari2015measuring}
Montanari F.,  Durrer R.,  2015, Journal of Cosmology and Astroparticle
  Physics, 2015, 070

\bibitem[\protect\citeauthoryear{Peebles}{Peebles}{1976}]{peebles76}
Peebles P.,  1976, \mn@doi [Astrophys. J.] {10.1086/154280}, 205, 318

\bibitem[\protect\citeauthoryear{Peebles}{Peebles}{1980}]{Peebles80}
Peebles P.,  1980, The Large-Scale Structure of the Universe.
Princeton Series in Physics, Princeton University Press, Princeton, NJ

\bibitem[\protect\citeauthoryear{{Planck Collaboration}}{{Planck
  Collaboration}}{2015}]{2015arXiv150201589P}
{Planck Collaboration} 2015, preprint, \href
  {http://adsabs.harvard.edu/abs/2015arXiv150201589P} {} (\mn@eprint {arXiv}
  {1502.01589})

\bibitem[\protect\citeauthoryear{Polarski \& Gannouji}{Polarski \&
  Gannouji}{2008}]{polarski08}
Polarski D.,  Gannouji R.,  2008, \mn@doi [Phys. Lett. B]
  {10.1016/j.physletb.2008.01.032}, 660, 439

\bibitem[\protect\citeauthoryear{Press}{Press}{2007}]{press2007numerical}
Press W.~H.,  2007, Numerical recipes 3rd edition: The art of scientific
  computing.
Cambridge university press

\bibitem[\protect\citeauthoryear{Raccanelli, Bertacca, Jeong, Neyrinck  \&
  Szalay}{Raccanelli et~al.}{2016}]{raccanelli2016doppler}
Raccanelli A.,  Bertacca D.,  Jeong D.,  Neyrinck M.~C.,   Szalay A.~S.,  2016,
  preprint (\mn@eprint {arXiv} {1602.03186})

\bibitem[\protect\citeauthoryear{Schlegel et~al.,}{Schlegel
  et~al.}{2011}]{schlegel2011bigboss}
Schlegel D.,  et~al., 2011, preprint (\mn@eprint {arXiv} {1106.1706})

\bibitem[\protect\citeauthoryear{Seo \& Eisenstein}{Seo \&
  Eisenstein}{2003}]{seo03}
Seo H.-J.,  Eisenstein D.,  2003, \mn@doi [Astrophys. J.] {10.1086/379122},
  \href {http://adsabs.harvard.edu/abs/2003ApJ...598..720S} {598, 720}

\bibitem[\protect\citeauthoryear{Tegmark}{Tegmark}{1997a}]{tegmark1997measure}
Tegmark M.,  1997a, Physical Review D, 55, 5895

\bibitem[\protect\citeauthoryear{Tegmark}{Tegmark}{1997b}]{tegmark1997measuring}
Tegmark M.,  1997b, Physical Review Letters, 79, 3806

\bibitem[\protect\citeauthoryear{Tegmark, Taylor  \& Heavens}{Tegmark
  et~al.}{1997}]{tegmark1997karhunen}
Tegmark M.,  Taylor A.~N.,   Heavens A.~F.,  1997, The Astrophysical Journal,
  480, 22

\bibitem[\protect\citeauthoryear{Trotta}{Trotta}{2008}]{trotta2008bayes}
Trotta R.,  2008, Contemporary Physics, 49, 71

\bibitem[\protect\citeauthoryear{Wang}{Wang}{2006}]{wang2006dark}
Wang Y.,  2006, The Astrophysical Journal, 647, 1

\bibitem[\protect\citeauthoryear{Wang}{Wang}{2012}]{wang2012robust}
Wang Y.,  2012, Monthly Notices of the Royal Astronomical Society, 423, 3631

\bibitem[\protect\citeauthoryear{Wang \& Steinhardt}{Wang \&
  Steinhardt}{1998}]{wang1998cluster}
Wang L.,  Steinhardt P.~J.,  1998, The Astrophysical Journal, 508, 483

\bibitem[\protect\citeauthoryear{Xu, Wang  \& Chen}{Xu
  et~al.}{2014}]{xu2014forecasts}
Xu Y.,  Wang X.,   Chen X.,  2014, The Astrophysical Journal, 798, 40

\bibitem[\protect\citeauthoryear{Yahya, Bull, Santos, Silva, Maartens, Okouma
  \& Bassett}{Yahya et~al.}{2015}]{yahya2015cosmological}
Yahya S.,  Bull P.,  Santos M.~G.,  Silva M.,  Maartens R.,  Okouma P.,
  Bassett B.,  2015, Monthly Notices of the Royal Astronomical Society, 450,
  2251

\bibitem[\protect\citeauthoryear{Yu \& Peebles}{Yu \&
  Peebles}{1969}]{yu1969superclusters}
Yu J.,  Peebles P.,  1969, The Astrophysical Journal, 158, 103

\bibitem[\protect\citeauthoryear{{eBOSS Collaboration}}{{eBOSS
  Collaboration}}{2016}]{eBOSSonline}
{eBOSS Collaboration} 2016, The extended {Baryon Oscillation Spectroscopic
  Survey} {(eBOSS)}, \url{http://www.sdss.org/surveys/eboss/}

\makeatother
\end{thebibliography}



\appendix

\section{FFT algorithm for a radial convolution} \label{sec:FFT-algorithm-for}

In this appendix we explain how to use a one-dimensional FFT algorithm
to solve a radial convolution integral like the one in Eq. (\ref{eq:convolved_spectra_integral}).
For the solution of the integral in Eq. (\ref{eq:convolved_spectra_integral}),
FFT algorithms are orders of magnitude faster and more precise in
comparison to numerical quadrature algorithms.

A radial convolution integral has the following expression:
\begin{equation}
c(x)=\int f(|\mathbf{x'}|)g(|\mathbf{x-x'}|)d^{3}x'.\label{eq:radial_convolution}
\end{equation}
The goal is to use a one-dimensional FFT algorithm to find the Fourier
transform and inverse transform of a radial function like $f(|\mathbf{x}|)$
and then use the convolution theorem. Thus, for any radial function
$f(|\mathbf{x}|)$, we have
\begin{equation}
\tilde{f}(k)=\frac{4\pi}{k}\int_{0}^{\infty}f(r)\mathrm{sin}(kr)rdr,
\end{equation}

\begin{equation}
f(r)=\frac{1}{(2\pi)^{3}}\frac{4\pi}{r}\int_{0}^{\infty}\tilde{f}(k)\mathrm{sin}(kr)kdk.
\end{equation}
Compared to the usual FFT methods, the sine functions are defined
on the whole positive real axis. Thus we need to assume to deal with
functions that are zero beyond some cutoff distance $\mathcal{R}$.
We also assume that $f(r+\mathcal{R})=f(r)$ and we discretize:
\begin{equation}
k_{l}=\frac{\pi}{\mathcal{R}}l\qquad\qquad r_{n}=\frac{\mathcal{R}}{\mathcal{N}}n
\end{equation}
with $n,l\in\{0,...,\mathcal{N}-1\}$. We proceed by defining $F_n\equiv r_{n}f(r_{n})$
and $\tilde{F}_{l}\equiv k_{l}\tilde{f}(k_{l})$. Then it can be
shown that:
\begin{equation}
\mathcal{S}(F_{n})\equiv\tilde{F_{l}}=2\sum_{n=0}^{\mathcal{N}-1}F_{n}\mathrm{sin}\left(\frac{\pi}{\mathcal{N}}nl\right)=-\mathrm{Im}\left[\sum_{n=0}^{M-1}\bar{F}_{n}\mathrm{exp}\left(-i\frac{2\pi}{M}nl\right)\right],
\end{equation}
\begin{equation}
\mathcal{S}^{-1}(\tilde{F_{l}})\equiv F_{n}=\frac{1}{\mathcal{N}}\sum_{l=0}^{\mathcal{N}-1}\tilde{F}_{l}\mathrm{sin}\left(\frac{\pi}{\mathcal{N}}nl\right)=-\frac{1}{2\mathcal{N}}\mathrm{Im}\left[\sum_{n=0}^{M-1}\bar{\tilde{F}}_{l}\mathrm{exp}\left(-i\frac{2\pi}{M}nl\right)\right],
\end{equation}
where we defined $M=2\mathcal{N}$ and the extended coefficients (similar
for $\bar{\tilde{F}}_{l}$):
\begin{equation}
\bar{F_{n}}\equiv\begin{cases}
F_{n} & 0\leq n<\mathcal{N}\\
0 & n=\mathcal{N}\\
-F_{2\mathcal{N}-n} & \mathcal{N}<n\leq2\mathcal{N}-1.
\end{cases}
\end{equation}
In this way for computing $\mathcal{S}$ and $\mathcal{S}^{-1}$ we
can just apply a normal 1D FFT algorithm on the new defined coefficients
and then take the imaginary part of it.

Considering all the coefficients and the cases with $k_{l}=0$ or
$n=0$, the final results for the radial and inverse transforms are:
\begin{equation}
\tilde{f}_{l}=\begin{cases}
\frac{\mathcal{R}}{\pi k_{l}}\mathcal{S}\left[f_{n}\frac{n\mathcal{R}}{\mathcal{N}}\right] & k_{l}\neq0\\
2\sum_{n=0}^{\mathcal{N}-1}f_{n}\left(\frac{n\mathcal{R}}{\mathcal{N}}\right)^{2} & k_{l}=0
\end{cases}\label{eq:radial_FFT}
\end{equation}
\begin{equation}
f_{n}=\begin{cases}
\frac{\mathcal{N}}{n\mathcal{R}}\mathcal{S}^{-1}\left[\tilde{f}_{l}\frac{\pi l}{\mathcal{R}}\right] & n\neq0\\
\frac{1}{\mathcal{N}}\sum_{l=0}^{\mathcal{N}-1}\tilde{f}_{l}\left(\frac{\pi l}{\mathcal{R}}\right)^{2} & n=0
\end{cases}\label{eq:inverse_radial_FFT}
\end{equation}
We can now use the formula (\ref{eq:radial_FFT}) to compute the discrete
Fourier transform of the radial functions $f(|\mathbf{x}|)$ and $g(|\mathbf{x}|)$.
Then we apply the convolution theorem to (\ref{eq:radial_convolution}),
i.e. $\tilde{c}(k)=\tilde{f}(k)\tilde{g}(k)$, and we use (\ref{eq:inverse_radial_FFT})
to find $c(x)$.

\section{The Alcock-Paczynski effect on the windowed spectra} \label{sec:The-AP-effect}

In this appendix we will study how to include the Alcock-Paczynski
effect on the modulus $k$, the cosine $\mu$, the volumes of the
shells in Eq. (\ref{eq:convolved_spectra_integral}) and the shell
radiuses $R_{i}$ in the window functions. To do so we need to decompose
$\boldsymbol{k}$ in $\boldsymbol{k_{||}}+\boldsymbol{k_{\bot}}$ along the line
of sight. By changing the cosmological parameters, the two components
of $\boldsymbol{k}$ transform as:
\begin{equation}
\mathbf{k}_{\perp}^{(2)}=\mathbf{k}_{\perp}^{(1)}\frac{D_{A}(z;\boldsymbol{\theta}_1)}{D_{A}(z;\boldsymbol{\theta}_2)}=\mathbf{k}_{\perp}^{(1)}\alpha(z),\qquad\qquad\mathbf{k}_{||}^{(2)}=\mathbf{k}_{||}^{(1)}\frac{H(z;\boldsymbol{\theta}_2)}{H(z;\boldsymbol{\theta}_1)}=\mathbf{k}_{||}^{(1)}\gamma(z),
\end{equation}
where we hid the dependence of $\alpha(z)$ and $\gamma(z)$ on the sets of cosmological parameters $\boldsymbol{\theta}_1$ and $\boldsymbol{\theta}_2$ for a simpler notation. The scale vector $\mathbf{x}$ will transform in similar way with
reciprocal quantities, so that the inner product $\boldsymbol{k\cdot x}$
is invariant under a parameter transformation. In the following we
will use the simplified notations:
\begin{equation}
\mathbf{k}_{2}=\mathbf{k}_{\perp}^{(1)}\alpha(z)+\mathbf{k}_{||}^{(1)}\gamma(z)\equiv\mathrm{U}(z)\,\mathbf{k}_{1},\qquad\qquad\quad\mathbf{x}_{2}\equiv\mathrm{U^{-1}}(z)\,\mathbf{x}_{1},
\end{equation}
where $\mathrm{U}(z)$ is a matrix depending on $z$. For volumes
and modules of the vectors we have:
\begin{equation}
d^{3}x_{2}=d^{3}x_{1}\frac{H(z;\boldsymbol{\theta}_1)D_{A}^{2}(z;\boldsymbol{\theta}_2)}{H(z;\boldsymbol{\theta}_2) D_{A}^{2}(z;\boldsymbol{\theta}_1)}\equiv d^{3}x_{1} \zeta(z),\qquad d^{3}k_{2}=\frac{d^{3}k_{1}}{\zeta(z)},
\end{equation}
\begin{equation}
k_{2}=\Upsilon(z)k_{1},\qquad\qquad x_{2}=\frac{x_{1}}{\Upsilon(z)},
\end{equation}
where $\Upsilon(z)$ was defined in Eq. (\ref{eq:AP_3}). We can now
include the previously mentioned AP effects in Eq. (\ref{eq:corr_spectra_ALL}).
We start with $i=j$ and $\boldsymbol{k=k'}$:
\begin{eqnarray}
\left[V_{i}\left\langle \delta_{\mathrm{U}_{i}\mathbf{k}}^{(i)}\delta_{-\mathrm{U}_{i}\mathbf{k}}^{(i)}\right\rangle \right]_{\mathrm{AP}} & = & \frac{\zeta_{i}V_{i}}{(2\pi)^{3}}\int d^{3}k'P(k')\tilde{W}_{i,AP}^{2}\left[\mathrm{U}_{i}\mathbf{k}-\mathrm{U}_{i}\mathbf{k}'\right]\label{eq:AP_corr_ii}\\
 & = & \frac{V_{i}}{(2\pi)^{3}}\int d^{3}k'P(\Upsilon_{i}k')\tilde{W}_{i,\mathrm{AP}}^{2}\left[\mathrm{U}_{i}\mathbf{k}-\mathrm{U}_{i}\mathbf{k}'\right],
\end{eqnarray}
where we changed the integral variable: $\mathbf{k'}_{\mathrm{old}}=\mathrm{U}_{i}\mathbf{k'}_{\mathrm{new}}$
so the Jacobian will be $d^{3}k'_{\mathrm{old}}=d^{3}k'_{\mathrm{new}}/\zeta_{i}$
and the integral is computed over all the space so the substitution
does not affect that. By defining the modified window function $W_{i,\mathrm{AP}}(x/\Upsilon_{i})\equiv W_{i}(x)$,
we note that the Fourier window function computed in a shifted vector
$\mathrm{U}_{i}\mathbf{k}$ is just $\tilde{W}_{i}\left(\mathbf{k}\right)$:
\begin{eqnarray}
\tilde{W}_{i,\mathrm{AP}}\left(\mathrm{U}_{i}\mathbf{k}\right) & = & \frac{1}{\zeta_{i}V_{i}}\int W_{i,\mathrm{AP}}(x)e^{i\,\mathrm{U}_{i}\mathbf{k}\cdot\mathbf{x}}d^{3}x\label{eq:window_AP}\\
 & = & \frac{1}{V_{i}}\int W_{i,\mathrm{AP}}(x/\Upsilon_{i})e^{i\mathbf{k}\cdot\mathbf{x}}d^{3}x\\
 & = & \tilde{W}_{i}\left(\mathbf{k}\right),
\end{eqnarray}
where we changed again variable in the second passage ($\mathbf{x}_{\mathrm{old}}=\mathrm{U}_{i}^{-1}\mathbf{x}_{\mathrm{new}}$).
By considering now the full convolved spectrum defined in Eq. (\ref{eq:convolved_spectra_integral}),
we can conclude that the full derivative with respect to the cosmological
parameters\footnote{The derivative formalized here needs to be computed to find the Fisher
matrix (Eq. \ref{eq:cFM_integral}).} is:
\begin{align}
\frac{d\mathrm{ln}\mathscr{P}_{ii}^{\mathrm{conv}}}{d\theta_{\alpha}}(k,\mu,z_{i};\boldsymbol{\theta})= & \frac{1}{\mathscr{P}_{ii}^{\mathrm{conv}}}G_{i}^{2}\frac{V_{i}}{(2\pi)^{3}}\left[\int d^{3}k'\frac{\partial P(k',z=0;\boldsymbol{\theta})}{\partial\theta_{\alpha}}\tilde{W}_{i}^{2}(|\mathbf{k-k'}|)\right.+\nonumber \\
 & +\left.\int d^{3}k'\frac{\partial P(k',z=0;\boldsymbol{\theta})}{\partial k'}\frac{\partial k'}{\partial\theta_{\alpha}}(k',\mu',z_{i};\boldsymbol{\theta})\tilde{W}_{i}^{2}(|\mathbf{k-k'}|)\right]+2\frac{\partial\mathrm{ln}G(z_i;\boldsymbol{\theta})}{\partial\theta_{\alpha}}.\label{eq:conv_der_param-1}
\end{align}
The integral in the second term\footnote{If we consider the cosmological parameters used for our computation
of the Fisher matrix (Eq. \ref{eq:FM_parameters}), then we note that
the first term of Eq. (\ref{eq:conv_der_param-1}) is not-zero for
$\theta_{\alpha}\in\{h,n_{s},\Omega_{\mathrm{b}}^{(0)},\Omega_{\mathrm{cdm}}^{(0)},w_{0}\}$,
instead the second and third ones are not-zero only for $\theta_{\alpha}\in\{\Omega_{\mathrm{b}}^{(0)},\Omega_{\mathrm{cdm}}^{(0)},w_{0}\}$.} is no longer a radial convolution. Thus we need a 3D FFT algorithm
to solve it, but for this specific integral this method is too memory-consuming.
Then we used a quadrature-algorithm to solve the integral. Note that
with a change of the integration variable $(k',\phi',\psi')$ such
that $\phi'$ becomes the angle between $\boldsymbol{k}$ and $\boldsymbol{k}'$,
the angular integral in $\psi'$ can be solved analytically.

The AP effect on the cross-correlation spectra leads to more difficulties.
For different bins $i\neq j$, we note that $\boldsymbol{k}$ and $\boldsymbol{k}'$
in Eq. (\ref{eq:corr_spectra_ALL}) change differently for a transformation
of the cosmological parameters. Thus we can not fix $\boldsymbol{k=k'}$
as we did in Eq. (\ref{eq:AP_corr_ii}) or (\ref{eq:corr_spectra})
because we would compute a correlation term between different modes
$\mathrm{U}_{i}\mathbf{k}$ and $\mathrm{U}_{j}\mathbf{k}$, which
have been neglected in this article. Then we rather fix the transformed
modes to be equal, i.e. $\mathrm{U}_{i}\mathbf{k}=\mathrm{U}_{j}\mathbf{k}'$,
and with similar steps we find:
\begin{eqnarray}
\left[\sqrt{V_{i}V_{j}}\left\langle \delta_{\mathrm{U}_{i}\mathbf{k}}^{(i)}\delta_{-\mathrm{U}_{i}\mathbf{k}}^{(j)}\right\rangle \right]_{\mathrm{AP}} & = & \frac{\sqrt{\zeta_{i}V_{i}\zeta_{j}V_{j}}}{(2\pi)^{3}}\int P(k')\tilde{W}_{i,AP}\left[\mathrm{U}_{i}\mathbf{k}-\mathbf{k}'\right]\tilde{W}_{j,AP}\left[\mathrm{U}_{i}\mathbf{k}-\mathbf{k}'\right]d^{3}k'\label{eq:corr_AP_ij}\\
 & = & \frac{\sqrt{\zeta_{i}V_{i}\zeta_{j}V_{j}}}{(2\pi)^{3}}\int P(k')\tilde{W}_{i}\left[\mathbf{k}-\mathrm{U}_{i}^{-1}\mathbf{k}'\right]\tilde{W}_{j}\left[\mathrm{U}_{i}\mathrm{U}_{j}^{-1}\left(\mathbf{k}-\mathrm{U}_{i}^{-1}\mathbf{k}'\right)\right]d^{3}k'\\
 & = & \frac{\sqrt{V_{i}V_{j}}}{(2\pi)^{3}}\sqrt{\frac{\zeta_{j}}{\zeta_{i}}}\int P(\Upsilon_{i}k')\tilde{W}_{i}\left[\mathbf{k}-\mathbf{k}'\right]\tilde{W}_{j}\left[\mathrm{U}_{i}\mathrm{U}_{j}^{-1}\left(\mathbf{k}-\mathbf{k}'\right)\right]d^{3}k',\label{eq:AP_final-1}
\end{eqnarray}
where again in the last passage we changed variable ($\mathbf{k'}_{\mathrm{old}}=\mathrm{U}_{i}\mathbf{k'}_{\mathrm{new}}$).
We see that this result generalizes the previous one given by Eq.
(\ref{eq:AP_corr_ii}) and (\ref{eq:window_AP}) for $i=j$, but at
the same way we could have computed the correlation at $\mathrm{U}_{j}\mathbf{k}$
and we would have found a different result for Eq. (\ref{eq:AP_final-1}).\footnote{This is equivalent to fix the condition $\mathrm{U}_{j}\mathbf{k}=\mathrm{U}_{i}\mathbf{k}'$.
Note indeed that if $\mathbf{k=k'}$, then the correlation $\left\langle \delta_{\mathbf{k}}^{(i)}\delta_{\mathbf{-k}}^{(j)}\right\rangle $
is symmetric for an exchange of the bins $i$ and $j$, but for $\mathbf{k\neq k'}$
then $\left\langle \delta_{\mathbf{k}}^{(i)}\delta_{\mathbf{-k'}}^{(j)}\right\rangle \neq\left\langle \delta_{\mathbf{k}}^{(j)}\delta_{\mathbf{-k'}}^{(i)}\right\rangle $
(see Eq. \ref{eq:corr_spectra_ALL}).}

To compute the plots in Fig. \ref{fig:ellipses_AP_effect} and \ref{fig:ellipses_AP_all}
and the data listed in Table \ref{table:percentage_difference}, we
have done the following approximations on the AP effect. Double-integrals
solved with quadrature-algorithms are time-consuming, thus for Euclid
we considered correlations only between adjacent bins. As show in
Sec. \ref{sub:Testing-the-assumptions} by the results without AP
effect, this represents a good approximation. For adjacent bins we
approximate $\mathrm{U}_{i}\mathbf{k}\thickapprox\mathrm{U}_{j}\mathbf{k}$,
thus this solves the problem described previously. Furthermore we
note that in Eq. (\ref{eq:AP_final-1}) it is no longer possible to
change the integration variable and solve analytically the integral
in $\psi'$. Quadrature-algorithms for triple integrals are too time-consuming,
then we approximate the factor $\mathrm{U}_{i}\mathrm{U}_{j}^{-1}\thickapprox\mathds{1}$
in Eq. (\ref{eq:AP_final-1}), i.e. we consider the same AP correction
on both the window functions $W_{i}(\mathbf{x})$ and $W_{j}(\mathbf{x})$.

We considered these approximations to show that the effects of the
window function and of the bin correlations remain similar when we
add the AP effect (Table \ref{table:percentage_difference_redshift}
and Fig. \ref{fig:ellipses_AP_effect}, \ref{fig:ellipses_AP_all}).
Nevertheless, the computations of the windowed spectra with the AP
effect are order of magnitudes slower compared to the ones computed
without it. Thus we conclude that the addition of the window function
or bins correlations does not increase computational times only without
including the AP effect.

\bsp	
\label{lastpage}
\end{document}